%% file: ms.tex
\documentclass[preprint,10pt]{article}

 \usepackage{graphicx, epstopdf}
\usepackage{color}
\usepackage[utf8]{inputenc}

\usepackage{amssymb, amsmath}
\usepackage{authblk}

\usepackage{caption}
\usepackage{multirow}
\usepackage{amsfonts}
\usepackage[math]{easyeqn}
\usepackage{algorithm,algpseudocode}
\usepackage{listings}

\include{def/commands}

\newcommand\mnewcommand[1]{%
\let#1\relax \newcommand#1 }

\newcommand{\code}[1]{{\ttfamily #1}}

\newcommand{\eff}{\mathit{eff}}
\newcommand{\power}{POWER9 cluster}
\newcommand{\mn}{MareNostrum}

\RequirePackage{amsfonts,amssymb,amsbsy,amsmath}

\begin{document}


  %
  \title{Heterogeneous CPU/GPU co-execution of CFD simulations on the POWER9
architecture: Application to airplane aerodynamics\footnote{© 2020 Elsevier.
This manuscript version is made available under the CC-BY-NC-ND 4.0 license
\texttt{http://creativecommons.org/licenses/by-nc-nd/4.0/} \newline
\texttt{https://doi.org/10.1016/j.future.2020.01.045}
}}

  
\author[1]{R. Borrell}
\author[1]{D. Dosimont}
\author[1]{M. Garcia-Gasulla}
\author[1]{G. Houzeaux}
\author[1]{O. Lehmkuhl} 
\author[2]{V. Mehta}
\author[1]{H. Owen}
\author[1]{M. V\'azquez}
\author[1]{G. Oyarzun\thanks{corresponding author: guillermo.oyarzun@bsc.es}} 

\affil[1]{Barcelona Supercomputing Center, c/ Jordi Girona 31 ,08034, Barcelona, Spain}
\affil[2]{NVIDIA, 2788 San Tomas Expressway, Santa Clara, CA 95051, USA}
  
\date{January 31, 2020}

  \maketitle
  \begin{abstract}
    High fidelity Computational Fluid Dynamics simulations are generally associated with large computing
    requirements, which are progressively acute with each new generation of
    supercomputers. However, significant research efforts are required to unlock the computing power of
    leading-edge systems, currently referred to as pre-Exascale systems, based on increasingly complex
    architectures. In this paper, we present the approach implemented in the computational mechanics
    code Alya. We describe in detail the  parallelization strategy implemented to fully exploit the
    different levels of parallelism, together with a novel co-execution method for the efficient
    utilization of heterogeneous CPU/GPU architectures. The latter is based on a multi-code co-execution
    approach with a dynamic load balancing mechanism. The assessment of the performance of all the
    proposed strategies has been carried out for airplane simulations on the POWER9 architecture
    accelerated with NVIDIA Volta V100 GPUs.


  \end{abstract}
  
  

%
\input{sections/01_introduction}

\input{sections/02_p9cluster}

\input{sections/03_application_context}

\input{sections/04_cpu_performance}
\input{sections/05_gpu_performance}

\input{sections/06_coexecution}
\input{sections/07_conclusions}

\section*{Acknowledgments}
This work is partially supported by the BSC-IBM Deep Learning Research Agreement, under JSA “Application porting, analysis 
and optimization for POWER and POWER AI”.
It has also been partially supported by the EXCELLERAT project funded by the European Commission's ICT activity of 
the H2020 Programme under grant agreement number: 823691.
It has also received funding from the European Union's Horizon 2020 research and innovation programme under grant agreement 
number: 846139 (Exa-FireFlows).
This paper expresses the opinions of the authors and not necessarily those of the European Commission. 
The European Commission is not liable for any use that may be made of the information contained in this paper.
This work has also been financially supported by the Ministerio de Economia, Industria y Competitividad, of Spain (TRA2017-88508-R).
The computing experiments of this paper have been performed on the resources of the Barcelona Supercomputing Center.


\bibliographystyle{is-unsrt}

\bibliography{ms}

\end{document}

%% file: def/commands.tex
\newcommand{\mybm}[1]    {\mbox{\boldmath{$#1$}}}

\newcommand{\boldo}    {{\mybm 0}}

\newcommand{\boldu}    {{\mybm u}}

\newcommand{\boldtau}  {{\mybm \tau}}

\newcommand{\defor}    {{\mybm \varepsilon}}

%% file: sections/01_introduction.tex
\section{Introduction}

On the road to exascale, supercomputers are becoming increasingly complex, the hybridization of their architectures being one of the most apparent trends. 
Heterogeneous computing has consolidated its position at the high end of the HPC market, basically in the form of GPUs introduced as co-processors. 
GPUs address some of the most significant exascale challenges such as the compute density and the energy costs.
For example, the Summit and Sierra supercomputers, based on the accelerated
POWER9 architecture, are placed on the first two positions of the
\textit{TOP500}~\cite{top500} list in November 2019.
Both supercomputers are also ranked in the fifth and tenth position of the contemporary \textit{Green500}~\cite{green500} list. 
In the $7th$ position of the same list there is the \mn{} \power{}, installed at the Barcelona Supercomputing Center, that we use in the present work.

The increasing complexity of the computing systems has its counterpart on the complexity of the software required for its efficient exploitation. 
A combination of various parallelization modes is necessary to harness all the levels of parallelism. 
Additionally, hardware heterogeneity poses new challenges such as the portability of the code across different devices or the balanced distribution of workload between them. 
In this paper, we describe the approach implemented in Alya to exploit a leading
edge heterogeneous architecture such as the \mn{} \power{}.
Alya~\cite{HOUZEAUX09,VAZQUEZ16b} is a high-performance 
computational mechanics code developed at the Computer Applications in Science and Engineering (CASE) department from BSC. 
It solves incompressible/compressible turbulent flows, solid mechanics, chemistry, particle transport, heat transfer, electrical propagation, etc. 
In this paper, we focus on airplane simulations that is one of its most computationally hungry applications and one of the research priorities at CASE. 
Actually, Alya can be obtained from the Unified European Applications Benchmark
Suite (UEABS)~\cite{ueabs} of the Partnership for Advanced Computing in Europe
(PRACE).

The contributions of this paper are twofold. 
Firstly we present a study of the performance 
of Alya on the two devices composing the \mn{} \power{}.
Secondly, we present a novel heterogeneous computing strategy, targeting the maximum occupancy of the entire node. 
The second objective is motivated by the performance assessment, from which we conclude that none of the node devices has a negligible contribution. 
We have investigated the concurrent use of both CPU and GPU to improve the efficiency of our application,  
referring to such form of heterogeneous computing as co-execution. 

General purpose programming models, such as OmpSs~\cite{ompss} or StarPU~\cite{augonnet2011starpu}, 
can also be employed for heterogeneous computing. Those are generally based on the taskification principles. 
For a Finite Elements code, a task can be defined as a step of the algorithm. Additionally, in order to expose more parallelism, a task 
can also be associated with a part of the mesh (a second level partition). Then,
a graph representing the dependencies among tasks is generated, so the 
scheduler can dynamically launch the executable tasks on the available resources. Note that for the best exploitation of this model each task
needs to be coded for all the available devices.
The trade-off between the optimal granularity for balancing and the associated scheduling and data transfer overheads needs to be considered as well. 
An example of this approach can be found in~\cite{Nesi19}.

A second option is to run each MPI process on a single device (either CPU or
GPU), adapting the domain partition to the relative performance of each device. 
This method has already been considered on heterogeneous systems~\cite{6975085,OYARZUN2018786}. 
Note that the partition adaptation can be based on performance measurements obtained during the execution, however, it does not act as a runtime mechanism that 
reacts instantaneously to any sort of hardware noise. 
Nonetheless, unlike taskification methods, once an optimized partition is settled no additional overhead remains, and  the repartitioning is not limited to the shared memory space.

In fact, both options are complementary since a low-level runtime mechanism should be more efficient, at overcoming unexpected imbalances, from a well-balanced partition than from an
unbalanced one. 
In this paper, we present a novel strategy for balancing the load of each MPI process according to the average performance of the device where it is executed. 
The underlying mesh partition is carried out with an in-house Space-Filling Curve (SFC) method~\cite{BORRELL2018}, 
to which we have incorporated partition correction coefficients to adapt the
partition to measurements obtained during the execution. 
The evaluation of these correction coefficients is based on the construction of a linear regression around each splitting point of the SFC. 
This strategy proves to be very robust even in heterogeneous systems and, as far as the authors of the paper know, it has not been proposed by any other author yet.
Other authors proved that in some application contexts an imbalanced execution
could outperform balanced ones~\cite{Lasto17,KHAL18}.  
However, in our examples, balancing the execution of the parallel processes
has always translated into a reduction of the time to solution.

The rest of the paper is organized as follows. 
In the next section, we present an overview of the hardware architecture and components of the MareNostrum POWER9 cluster. 
In Section~\ref{sec:app}, we present a top-down view of the background of this work: 
from the application problem under consideration (airplane simulations), through the physical and numerical models, 
to the implementation and parallelization strategies used in Alya. 
In Sections~\ref{sec:cpu} and~\ref{sec:gpu}, we assess the performance of Alya in the CPU and GPU devices composing the cluster, respectively. 
Then, in Section~\ref{partition}, we describe our load balancing strategy, which is the main building block of our co-execution algorithm, 
and we show its efficiency in the CPU and GPU devices separately. 
Afterwards, in Section~\ref{sec:coex}, we present our co-execution approach and
assess its performance for an airplane simulation on a 176 million element mesh. 
Finally, we summarize our contributions in Section~\ref{sec:conclus}.

%% file: sections/02_p9cluster.tex
\section{MareNostrum \power}
\label{sec:cluster}

In this paper we evaluate the performance of the MareNostrum \power{} accelerated with Volta V100 GPUs. 
The peak performance of the \power{} is over 1.5 Petaflop/s. 
The cluster consists of 2 login nodes and 52 compute nodes POWER9 AC922 with 2 sockets each, 
512 GB of main memory and 4 Volta V100 accelerators (16 GB discrete
memory each). 
All the 52 nodes are connected via Mellanox EDR interconnect fabric.

Following the current trend in HPC node design, the POWER9 nodes are high throughput nodes. 
Each compute node has 2 x POWER9 8335 @3.0 GHz with 20 physical cores each. 
With 4 SMT threads per core, each node can run up to 160 CPU threads. 
The 512 GB of main memory is distributed in 16 DIMMS of 32 GB each operating at 2666MHz. 
The 4 Volta V100 accelerators have a 7.8 TFLOPS double precision peak performance giving each one a total of 30.8 TFLOPS. 
The  POWER9 nodes include high throughput NVLINK 2.0 connectors. 
Each GPU has 6 NVLINK connectors, which are attached to the neighboring GPU as well as CPU, 
giving an aggregate bandwidth of 150 GB/s for GPU to GPU as well as GPU to CPU communication.
Details on the node architecture can be found in ~\cite{power9}.

%% file: sections/03_application_context.tex
\section{Application context}
\label{sec:app}

\subsection{Motivation}

Around half of the energy spent worldwide in transport activities is dissipated by the turbulent motion in the immediate vicinity of solid surfaces. 
Consequently, the objectives set by the Advisory Council for Aeronautics Research in Europe, in terms of fuel consumption and noise emission, 
have prompted a large number of research activities for innovative solutions to get more ecologic and more economic airplanes.

One of the proposed activities aims at obtaining accurate simulations of airplanes, 
to predict and understand the aerodynamics as a whole, 
as well as the interactions between the different airplane components (fuselage, wings, nacelles, etc.). 
Traditional Reynolds averaged Navier-Stokes (RANS) approaches have demonstrated reasonable success in the prediction of integral quantities of interest 
(e.g., lift, drag) in external aerodynamics at low angles of attack, 
where the boundary layers remain largely attached~\cite{Tinoco}. 
Maximum lift conditions (where the boundary layer may be separated in a limited region) or fully stalled regimes 
(where the boundary layer has undergone massive separation) have proven to be challenging to predict with RANS closures~\cite{Slotnick}, 
particularly for steady RANS approaches as the underlying flow field possesses large-scale and low-frequency unsteadiness.

To guarantee accurate results in such high lift configurations, high-fidelity simulations based on Large Eddy Simulation (LES) models are necessary. 
These simulations are transient, with relatively small time steps and require very fine meshes. 
Harness the potential of leading-edge HPC systems is thus a must to carry out such simulations. 
The present work aims at preparing Alya for the efficient exploitation a modern system such as the POWER9 accelerated architecture, 
demonstrating its readiness for accurate airplane simulation using a state of the art LES.

\subsection{Physical and numerical model}
\label{sec:numer}

To solve the aerodynamics of the aircraft configuration,
we consider the spatially filtered Navier-Stokes equations which model turbulent incompressible
flows \cite{batchelor_2000}. The incompressible flow assumption is valid in the present case as
the inflow conditions reproduce a landing configuration at a Mach number of 0.172. 
Let $\rho$ and $\mu$ be the density and viscosity of the fluid, respectively.
The problem statement is: find the filtered fluid velocity $\boldu$ and modified pressure $p$ in a domain $\Omega$ and during a given time interval such that
\begin{EQA}[l]
  \displaystyle{\rho \frac{\partial \boldu}{\partial t}}
   + \rho \left[ 
  2\mathbf{u} \cdot \defor(\mathbf{u})
  +\left(\nabla\cdot\mathbf{u}\right)\mathbf{u}
  -\frac{1}{2} \nabla |\mathbf{u}|^2 \right] \\
   \qquad \qquad \qquad \qquad \qquad \qquad
  - \nabla \cdot[ 2 \mu \defor(\boldu) ]
  + \nabla p + \nabla \cdot \boldtau = \boldo, \label{eq:momentum}\\
   \qquad\qquad\qquad 
   \nabla \cdot \boldu = 0,
   \label{eq:continuity}
\end{EQA}
together with initial and boundary conditions. 
The velocity strain rate is defined as
$\defor(\boldu) := \frac{1}{2} ( \nabla \boldu + \nabla \boldu^t)$, and $\boldtau$ is the subgrid
scale stress tensor \cite{Pope:1346971}. 
At the continuous level, the convective term, between brackets in the equation, can be written in many equivalent ways.
In this work, we consider the EMAC formulation~\cite{lowdis} which, at the
discrete level, ensures stability and conserves energy as well as linear and angular momentum.

To model the subgrid scale stress tensor, Vreman LES model is used~\cite{VRE04b-A}. 
Wall modeling is based on a law of the wall, extensively described in ~\cite{wall}. 
Space discretization is carried out using the Galerkin Finite Elements method \cite{j2005introduction}. 
We use a fractional step method together with a Runge-Kutta scheme of third order as a time discretization scheme. 
Momentum is advanced explicitly, while the projection step of the algorithm, 
which consists in solving a Poisson equation for the pressure, provides the pressure stabilization \cite{codina}. 
This associated algebraic system is solved with the Conjugate Gradient (CG) method with Jacobi diagonal scaling as preconditioner~\cite{YSaad03}. 
Details on the numerical and time integration schemes are provided in ~\cite{lowdis}.

\begin{algorithm}[h!]
  \begin{algorithmic}[1]
  \State Element assembly the Laplacian matrix.
  \For   {Time steps}
  \For   {Runge-Kutta steps}  
    \State {\bf  Element assembly}: assembly of momentum equations.  \label{ite:rk1}
    \State {\bf  Boundary assembly}: assembly of wall model.         \label{ite:rk2}
    \State Momentum update.                                          \label{ite:rk3}
    \State Velocity correction.                                      \label{ite:rk4}
  \EndFor
    \State {\bf  Algebraic solver}: solution of pressure equation with the preconditioned CG. \label{ite:press}
  \EndFor
\end{algorithmic}
\caption{Main steps of the fractional step method.}
\label{alg:kernels} 
\end{algorithm} 

We outline the main steps of the algorithm for the integration of the above equations over time in  
Algorithm \ref{alg:kernels}. 
At each time step, three Runge-Kutta steps (\ref{ite:rk1}-\ref{ite:rk4}) are first performed. 
For each of these Runge-Kutta steps, element and boundary loops assemble the momentum equations and the law of the wall, respectively. 
Then, the preconditioned CG method is used to solve the pressure Poisson equation in step~\ref{ite:press}. 
As will be shown in the performance analysis section, the most time consuming  kernel for the test case considered in this paper is the elements assembly. 

\subsection{Test case: airplane simulation}
\label{sec:test_case}

The selected airplane configuration used in the present study is the JAXA Standard Model (JSM)~\cite{Yokokawa2} which is a wing-body high lift system. 
The geometry used is the one recommended in the 3rd AIAA CFD High Lift Prediction Workshop~\cite{nasa}, 
where the JSM is studied in a nominal landing configuration (single segment baseline slat and single segment $30$ degree flap) with support brackets in, 
without nacelle/pylon and in free stream conditions. The Mach number for these conditions is 0.172, 
with a Reynolds number based on the mean aerodynamic chord of $1.93$ M and an angle of attack of $18.58$ degrees.
At both the inlet and the boundaries that are far from the region of interest,
an homogeneous velocity is prescribed. For the aircraft walls an equilibrium
wall model is used in ~\cite{wall}. For the initial condition, a uniform velocity is used in the whole domain except on the aircraft walls where the 
velocity is modified such that its normal component is zero.  

Two meshes have been generated to carry out the experiments presented in this paper. 
On the one hand, a fine mesh designed for the production simulations. 
On the other hand, a coarse mesh for the battery of computational and efficiency tests. 
We have taken care that during these computational tests, the load per computing node is similar to that obtained when running the fine mesh simulations.

\begin{figure}[h!tbp]
  \centering
  \includegraphics[width=0.80\textwidth]{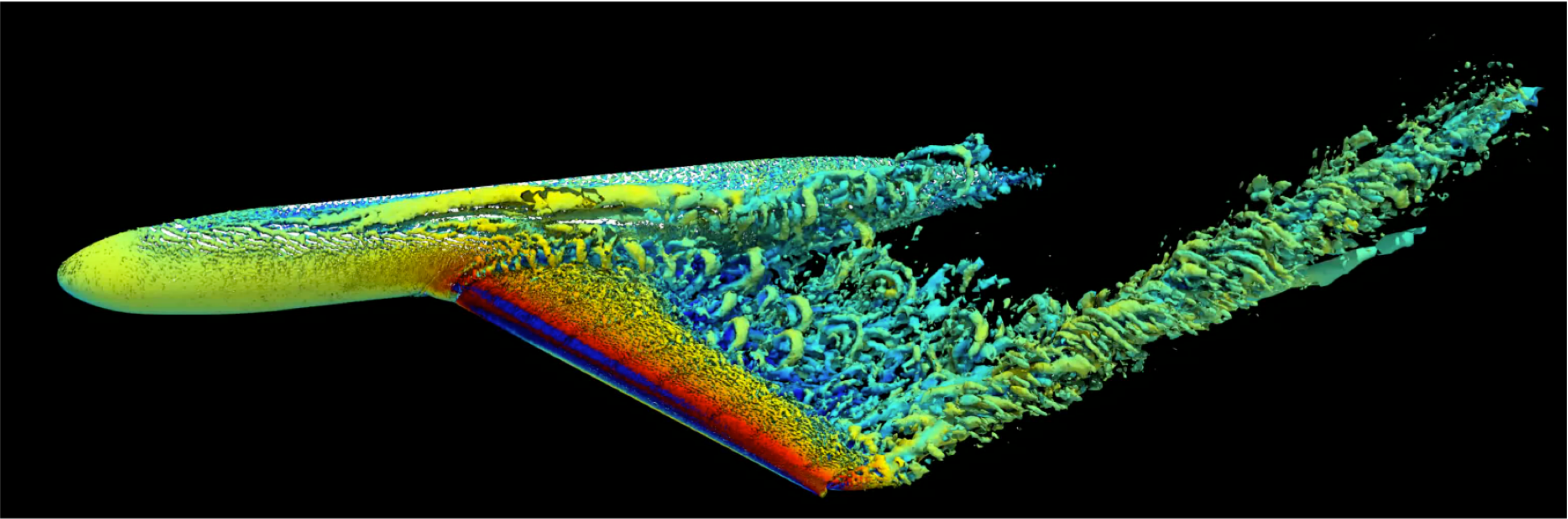}
  \caption{Snapshots of the Q-criterion.
  }
  \label{fig:qcriterion} 
\end{figure}

The fine mesh is a linear hybrid mesh consisting of $176$M elements (with
tetrahedron, prisms, and pyramids). 
It has approximately three elements in the inner region of the boundary layer
and about $10$ elements in the outer layer region. 
Similar mesh resolutions have been used by the authors to simulate similar flows with success~\cite{lehmkuhl}. 
Figure~\ref{fig:qcriterion} depicts an instantaneous snapshot of the Q-criterion.
The coarse mesh has similar characteristics but is composed of $31.5$M elements.

\subsection{Code and Parallelization}

We have implemented all the computational strategies presented in this paper in the simulation code Alya. 
Alya is a high-performance computational code for multi-physics simulations, developed mainly at the Barcelona Supercomputing Center. 
It is written in Fortran and includes different levels of parallelization and optimization to scale on a variety of supercomputer architectures.

We can divide the parallelization strategy implemented in Alya in three categories: 
distributed memory for inter-node parallelism; shared memory for intra-node parallelism; 
SIMD (Single Instruction Multiple Data) and SIMT (Single Instruction Multiple Thread) for CPU and GPU, respectively. 

\begin{figure}[h!tbp]
  \centering
  \includegraphics[width=0.99\textwidth]{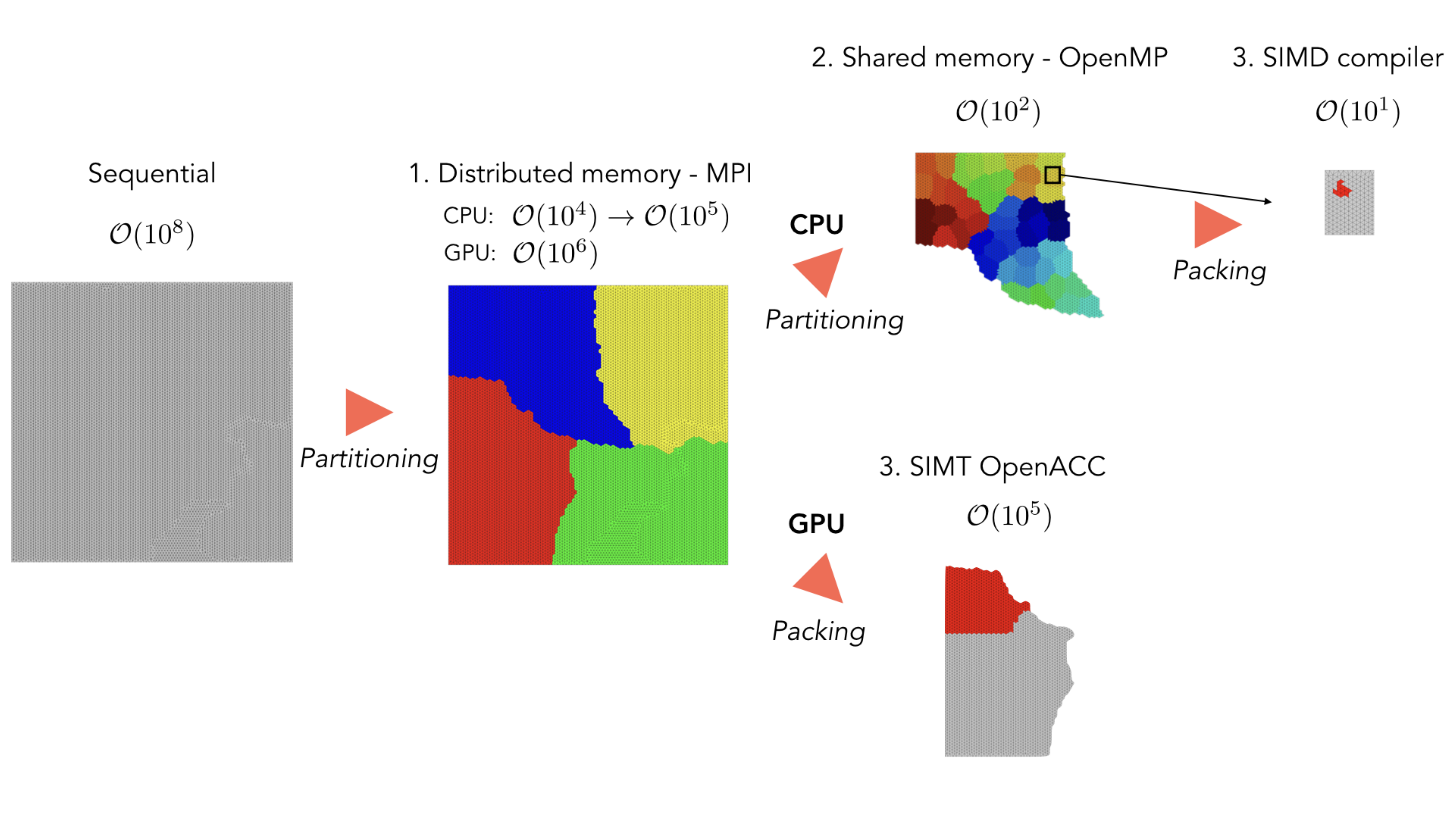}
  \caption{Three levels of parallelism used for element assembly, for CPU (top) and GPU (bottom).
  The orders of magnitude of the numbers of elements involved at each parallelization level are
  shown in parenthesis.}
  \label{fig:parall} 
\end{figure}

Figure~\ref{fig:parall} illustrates these parallelization levels for the specific case of the element
and boundary assemblies, 
where the routes to CPU and GPU optimizations are distinguished (top and bottom, respectively).
For both routes, a partitioning is first carried out where subdomains are attached to MPI processes;
the size, in terms of elements, of the subdomains of this first level ranges from $10^4$ to $10^5$ for the CPU and 
 $10^6$ for the GPU. 
As far as CPU is concerned, a subsequent partitioning of each MPI subdomain
is performed for a second level of parallelism; here, task parallelism will be used on subsets of
the order of $10^2$ elements. Finally, a packing is done for
SIMD parallelism. The packing consists in grouping elements for vectorization; the size
of the packs ranges from 8 to 32 according to the CPU employed. 
As far as GPU is concerned, the
packing is directly carried out on the MPI subdomains. In this case, the size of the pack is of the order of
$10^5$ elements that are concurrently executed on the GPU through stream processing. 

A typical pack for both CPU and GPU is colored in red in the figure.
In the case of co-execution, both routes coexist in the same MPI environment: some MPI subdomains
will be treated for CPU and the remaining for GPU. In this case, the solver
is only executed on the GPU, using the same partition generated for
the assembly and linking each MPI process with one GPU. The following sections explain in detail these different parallelization levels.

\subsubsection{Distributed Memory}

The distributed memory parallelization in Alya is achieved via a classical substructuring technique extensively described 
in~\cite{HOUZEAUX09,VAZQUEZ16b}. The mesh is partitioned into disjoint sets of elements composing subdomains by means of an 
in-house SFC-based method~\cite{BORRELL2018}, and  each set is assigned to a MPI process. We duplicate nodes located in the interface between neighboring 
subdomains; we refer to them as interface nodes. In the Finite Elements context, the element and boundary assemblies do not require halo elements and do not involve communication.
Point-to-point communications take place mainly in the sparse-matrix vector products (SpMV) present in the iterative solver: 
exchange of the SpMV result is carried out at the interface nodes. 
The iterative solver also involves collective reductions whenever a dot-product is computed. 
The iterative CG solver requires one SpMV and two dot-products per iteration.

The mesh partition step is crucial to achieving good performance. 
On the one hand, a convenient mesh partition should provide a well-balanced workload distribution among the processes,
but also minimize the communications during SpMV operations, 
which we achieve by reducing the number of interface nodes. 
On the other hand, any kind of communication implies a synchronization between the MPI processes. 
This synchronization can have serious implications on the performance if the workload distribution between MPI processes is not well balanced. 
We present further details on the mesh partitioning and load balancing strategy implemented in Alya in Section~\ref{partition}.
Note that many MPI processes can send their work to the same GPU without losing
any coherence in the results. In such cases, the NVIDIA Multi-Process Service (MPS) is used to manage and schedule the work coming from the different MPI processes.
\subsubsection{Shared Memory} 

We implement the shared memory parallelization using OpenMP, and mainly a loop based
parallelism approach (e.g., in each of the kernels of the iterative solver). 
OpenMP threads share the memory space, therefore the communication between them does not need to be explicit and load imbalance is easy to address because data is shared.

Although we mentioned that we use loop parallelism all through the code, in the case of the element and boundary assemblies, 
we consider a task-based parallelization instead, as it brings significant benefits in term of performance~\cite{MPI_x}. 
The task-based parallelization leverages two new features of the OpenMP 5.0 standard~\cite{omp5}: \texttt{mutexinoutset} and \texttt{iterators}. 
These features are not available yet in any of the compilers and runtimes that
we are using (GFortran, XLF and PGIF90). 
For this reason, in our hybrid experiments, we will use OmpSs~\cite{ompss}. 
OmpSs is a parallel runtime we use as a forerunner for new features of OpenMP.

In the case of task parallelism, each OpenMP task executes the assembly of an OpenMP subdomain. 
These subdomains are constructed by partitioning each MPI subdomain, such that a typical size of the OpenMP subdomain is ${\cal O}(10^2)$ elements, 
to obtain optimum speedup~\cite{MPI_x}. 
Figure \ref{fig:parall} (top) illustrates this repartitioning used for the task parallelism of the element assembly. 
We employ a similar strategy for the boundary assembly.

\subsubsection{SIMD and stream processing}
\label{simdsimt}
This last level of parallelism consists of reorganizing the data and creating
data structures that facilitate SIMD execution (CPU) and SIMT (GPU). In the GPU
execution model, two additional parameters are needed: threads and blocks. The workload is
 distributed into thousands of threads that are grouped into blocks. Each block contains the same
 amount of threads. The threads within the block are executed in SIMD mode, so at this level CPU
 and GPU optimal data structures are the same. The GPU manages the number of blocks that can keep
 active depending on the memory requirements of the threads (i.e. number of registers,
 shared memory and number of threads per block).

The element assembly and iterative solver require two different treatments since its algorithms have different memory accesses and arithmetic intensities. 
The former is tightly coupled with the domain representation, thus requiring special attention for mapping efficiently its indirect memory accesses. 
On the other hand, the latter consists of basic algebraic operations commonly found in libraries that are already highly optimized for most of the architectures.

\paragraph{Element and boundary assemblies}

A typical implementation of an element assembly (similarly a boundary assembly) consists of a loop over each single element and
for each element are carried out the following three steps:
\begin{enumerate}
\item Gather: global mesh arrays are gathered into element arrays.
\item Computation: computations of all the terms of Equation \eqref{eq:momentum} in element arrays.
\item Scatter: the element arrays are scattered into global arrays.
\end{enumerate}

We generate a new data structure to enhance the efficiency of the assembly. 
First, we carry out a data reordering to store contiguously in memory the elements that share the same kind (tetrahedron, hexahedron, prism, and pyramid) 
and the same integration rule, thus defining different categories. 
Then, we group such elements into packs of size {\tt PACK\_SIZE}, each pack containing {\tt PACK\_SIZE} elements of the same category. 
Note that zeros are padded in the data structure when elements of the same category are not enough to fill a pack. 
Finally, the assembly runs on this pack of elements instead of on every single element.

This approach has a two-fold benefit. On the one hand, it improves data locality, because it stores elements in dense packs. 
On the other hand, the code exposes the SIMD/SIMT potential to the compiler.

We can achieve a common implementation for CPUs and GPUs. This is illustrated for the assembly of the mass matrix. 
We denote {\tt nnode} and {\tt ngaus} as the numbers of nodes and Gauss integration points of an element. 
{\tt Ae}, {\tt J}, {\tt w} and {\tt N} are the element matrix, the Jacobian, the weight of the integration point and the shape function, respectively. 
In a Fortran code, the computation of one pack of the element mass matrix can be written as shown in Listing \ref{lst:code-mass-matrix}.

In a classical approach, we note that the first index {\tt PACK\_SIZE} would not exist as elements are assembled one by one.
In our implementation, we observe that if OpenACC is not defined, then the innermost dimension of the arrays is the pack of elements used for CPU. 
On the other hand, by defining OpenACC, the loop is explicitly unrolled for {\tt ipack}, from 1 to {\tt PACK\_SIZE}, thus enabling the parallelization with OpenACC.
Despite the CPU and GPU implementations are compiled with different {\tt
PACK\_SIZE} and compilation flags, the codes can interact with each other by
using the co-execution model explained in Section~\ref{sec:coex}

\begin{minipage}{\linewidth} 
\begin{lstlisting}[frame=single,label=lst:code-mass-matrix,caption={Typical Finite Elements calculation: mass matrix}]
#ifdef OPENACC
#define DEF_PACK ipack
  use openacc
#else 
#define DEF_PACK 1:PACK_SIZE    
#endif

#ifdef OPENACC
!$acc data create(...)  &
!$acc copyin(...)
!$acc parallel loop gang vector default(present) 
do ipack = 1,PACK_SIZE                        
#endif

  do ig = 1,ngaus
     do jn = 1,nnode
        do in = 1,nnode
           Ae(DEF_PACK,in,jn) = Ae(DEF_PACK,in,jn) + J(DEF_PACK,ig) &
                * w(DEF_PACK,ig) * N(DEF_PACK,in,ig) * N(DEF_PACK,jn,ig)
        end do
     end do
  end do

#ifdef OPENACC
end do
!$acc end parallel loop
!$acc end data
#endif
\end{lstlisting} 
\end{minipage}

\begin{figure}
\centering
  \includegraphics[width=.55\textwidth]{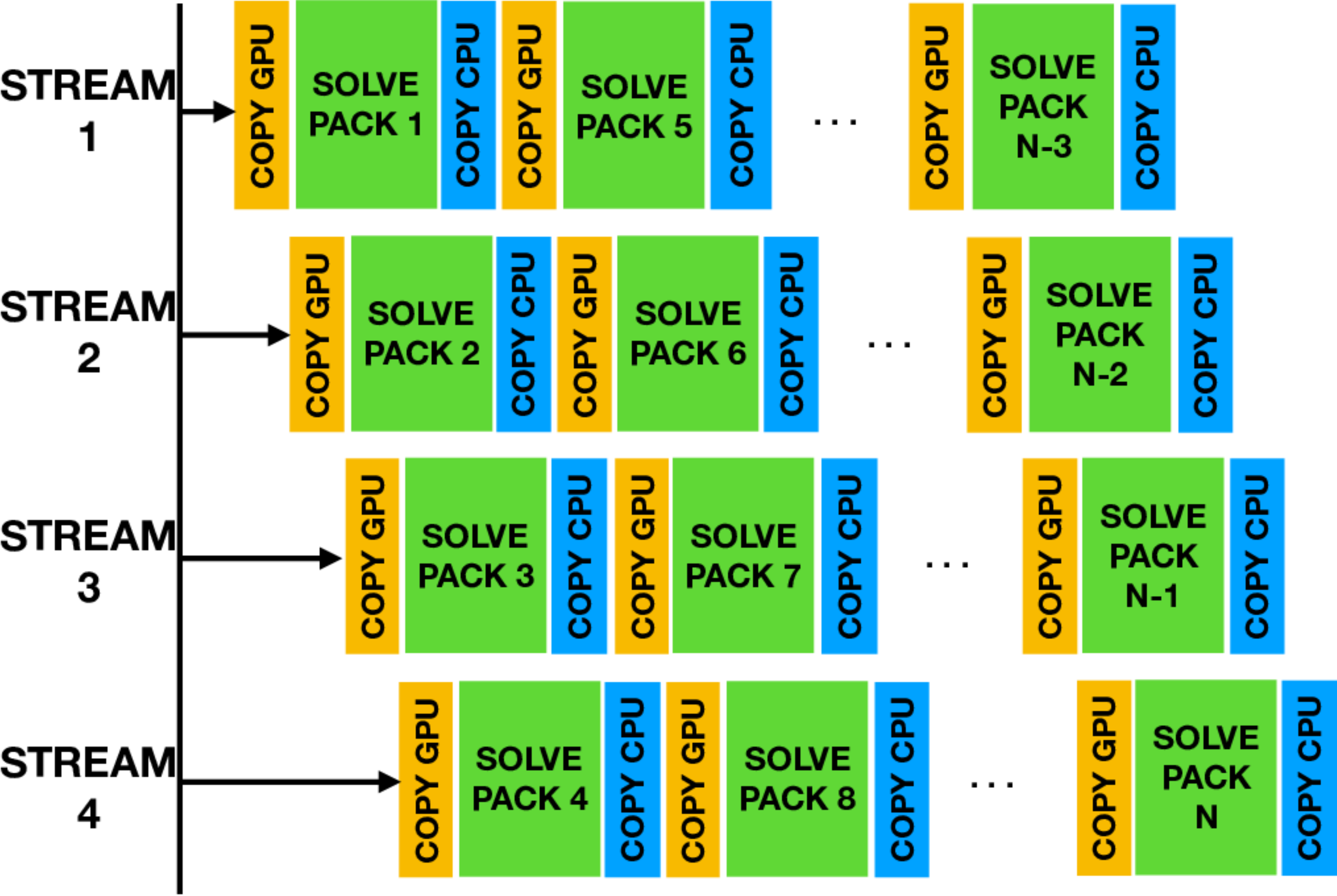}
   \caption{Pipeline Execution model.} 
   \label{fig:gpupipeline}
\end{figure}

In the GPU implementation, the {\tt PACK\_SIZE} defines the total number of threads launched by each kernel call, and consequently, 
determines the number of kernel instances in each parallelized loop ({\tt PACK\_SIZE} = blocks$\times$threads ). 
The performance of the kernels depends on the characteristics of the grid execution (threads and blocks), 
and in Alya's case it is set by using the \texttt{vector\_length} clause from
OpenACC for selecting the number of threads that run in SIMD mode. Note that
commonly there are many packs. In the GPU execution, each of them becomes a kernel call.  
In production cases, the solution of the element assembly is more complicated
than the mass matrix calculation, demanding to keep many arrays uploaded into
the GPU memory. In our code, we have adopted an explicit management of
the memory transfers to have more control in the communication schemes that involve
overlapping techniques. As consequence of this memory management, our implementation is limited to work
 within the boundaries of the physical memory of the GPU (16GB).  
 The grouping strategy allows to batch the data in packs that are
processed sequentially and do not overflow the GPU memory.
Since the kernel executions are independent to each other, 
then a pipeline execution model is suitable for overlapping data transfers and
computations and increasing the GPU occupancy (see Figure~\ref{fig:gpupipeline}). 
The concurrent execution is obtained by launching of multiple asynchronous streams using the clause \texttt{async()} in the  OpenACC loop directives.

In this implementation, one of the factors affecting the most the performance is the pack size.
We will discuss more in details the optimum size and how it differs when using CPUs or GPUs in the following sections. 
As we show in Sections~\ref{sec:opt_pack_cpu} and~\ref{sec:gpu}, a typical pack size for CPU is of the order of ${\cal O}(10^1)$ elements, 
while it is of the order of ${\cal O}(10^5)$ for GPU.

\paragraph{Iterative solver}

In the present work, we execute the CG iterative solver exclusively on GPUs. 
We build the GPU solver library of Alya on top of NVIDIA’s high-performance cuBLAS, cuSparse and cuSolver. 
The library also includes CUDA kernels, for functionalities not available in CUDA libraries. 
The heart of the library is composed of a sparse matrix-vector product, which is implemented based on cuSparse as well as native CUDA kernel. 
The native CUDA kernel performs better in case of Alya matrices, as we can tune it according to the bandwidth of the matrices. 
In the distributed-matrix vector product, we overlap the communications required at the subdomains interface, with the computation of the product for the inner nodes.

Alya also provides a pool allocator for managing GPU memory. 
The pool allocator reduces the overhead of both memory allocation and release. 
It also enables multiple solvers to operate together on different components of the linear system. 
Alya provides interfaces to CUDA libraries via C functions, and the solvers are natively written in Fortran, enabling us to add easily new solvers and preconditioners.

\subsection{Performance characterization}
\label{sec:perf_char}

As a first step in the performance analysis, we have executed the airplane test case of 31.5M (coarse mesh) on 320 MPI processes using 4 nodes of the \power, 
running only on the CPUs. 
For this analysis we have used Extrae~\cite{extrae} to obtain a trace of the execution and Paraver~\cite{paraver} to visualize it (the overhead introduced by the tracing tools is under 4\%~\cite{alyaAdvancedPASC20}).

\begin{figure}[h!tbp]
  \centering
  \includegraphics[width=\textwidth]{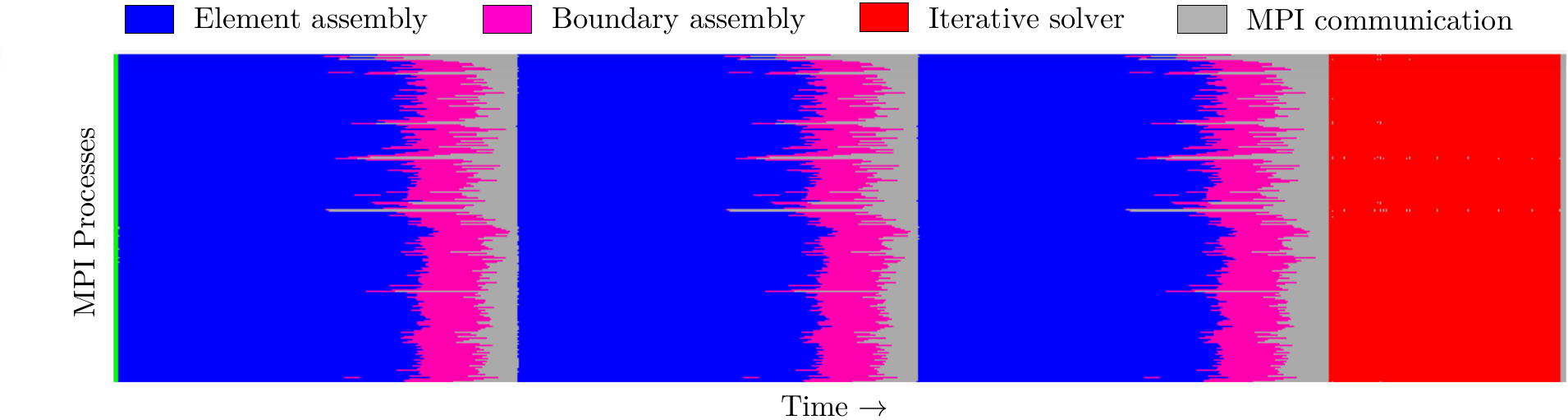}
  \caption{Trace of one time step of the Airplane simulation showing the different phases of computation.}
  \label{fig:trace_phases} 
\end{figure}

Figure~\ref{fig:trace_phases} shows a timeline of a real execution of the airplane simulation. 
The X-axis represents the time, and the Y-axis represents the different MPI processes. 
The different colors correspond to the three main phases of a time step: 
the element assembly (blue), the boundary assembly (pink), and the iterative solver (red). 
The gray color represents the idle time due to the load imbalance or MPI communications. 
The three consecutive element and boundary assemblies correspond to the three
steps of the Runge-Kutta algorithm mentioned in Section \ref{sec:numer}.

\begin{figure}[h!tbp]
  \centering
  \includegraphics[width=\textwidth]{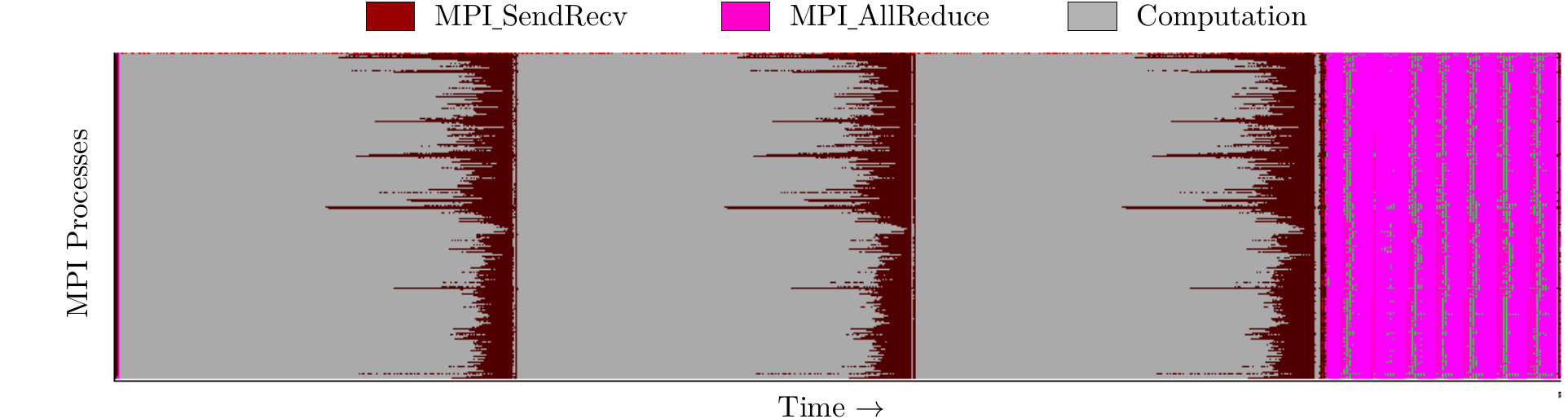}
  \caption{Trace of one time step of the Airplane simulation showing the MPI calls.}
  \label{fig:trace_mpi} 
\end{figure}

Figure~\ref{fig:trace_mpi} shows the different MPI calls invoked during one time step. 
In this case, the colors correspond to the MPI calls executed while the gray color corresponds to useful computation. 
We can observe the collective communication (MPI\_AllReduce) in the algebraic solver and the point-to-point communication (MPI\_SendRecv) 
after the boundary assembly and before the next element assembly: 
this communication corresponds to the residual momentum exchange at the interface nodes, required by the Runge-Kutta integration scheme.

\begin{table}[htbp]
 \caption{Counters and metrics for each kernel (LB: Load Balance, FP: Floating Point).}
\centering
\small{
\begin{tabular}[c]{l rrr}
\hline
                        & \textbf{Element}     & \textbf{Boundary}   & \textbf{Solver}\\\hline
\textbf{Arith.Int.}     & 0.05                 & 0.01                & 0.02\\\hline
\textbf{FP Ops.}         & 375.69$\times 10^9$  & 8.27$\times 10^9$   & 18.14$\times 10^9$\\
\textbf{\%FP Ops.}       & 93.43\%              & 2.06\%              & 4.51\%\\
\textbf{LB FP Ops.}      & 0.90                 & 0.68                & 0.85\\\hline
\textbf{Load+Stores}    & 1.03$\times 10^{12}$  & 120.68$\times 10^9$ & 113.62$\times 10^9$\\
\textbf{\% Load+Stores} & 81.47\%              & 9.55\%              & 8.99\%\\
\textbf{LB Load+Stores} & 0.92                 & 0.69                & 0.86\\\hline
\textbf{Time [s]}       & 589.17               & 111.25              & 83.28\\
\textbf{\% Time}        & 75.18\%              & 14.20\%             & 10.63\%\\
\textbf{LB Time}        & 0.92                 & 0.73                & 0.82\\
\hline
\end{tabular}
}
\label{tab:counters}
\end{table}

In the trace, we also collect PAPI performance counters, from which we compute the arithmetic intensity $I$. 
The arithmetic intensity is a common metric in the HPC community. 
It characterizes a computation phase, kernel or application with no (or very little) dependence on the hardware platform where it will run.

The arithmetic intensity $I$ is defined as $ I = \frac{W}{D} $, where $W$ is the computational work, in our case measured
 as the number of double precision floating point operations performed as
reported by hardware counters, and $D$ is the data moved measured in bytes. 

In Table~\ref{tab:counters} we can see different metrics for each of the computational kernels involved in the simulation. 
The first row contains the arithmetic intensity computed as explained above. 
We can see that the arithmetic intensity is higher in the element assembly than in the other phases.

Rows \textit{FP Ops.} and \textit{Load+Stores} contain the total number of instructions executed of each type. 
In the following rows (\textit{\%FP Ops.} and \textit{\% Load+Stores}) we see respectively the percentage of this kind of instructions that correspond to each phase.
Finally \textit{LB FP Ops.} and \textit{LB Load+Stores} correspond to the load balance between the different processes for each hardware counter.

In the last rows we see the timing metrics for each kernel; 
the elapsed time per step, the percentage of time that corresponds to each kernel of the total elapsed time and the load balance.
We remark that the element assembly performs 93.43\% of the floating point operations, while it only accounts for 75.18\% of the execution time. 
This observation suggests that the element assembly kernel is more efficient than the other ones, assuming that the processor is performing floating point operations most of the computing time.

For this study and all the remaining ones, we will compute the load balancing
(LB) as:
\begin{EQ}[rcl]
 LB &=& \frac{average}{max}.
\end{EQ}

%% file: sections/04_cpu_performance.tex
\section{CPU performance analysis}
\label{sec:cpu}

In this section, we evaluate the performance of the CPUs of the cluster using the airplane test case presented 
in Section~\ref{sec:test_case} for the coarser mesh of 31.5M elements. As we have seen in Section~\ref{sec:cluster}, each node of the \power{} is composed of two sockets, each one containing 20 cores. 
Each core can run up to 4 SMT threads, and each pair of cores share L2 and L3 caches.

\begin{table}[htbp]
 \caption{Compilers used for experiments.}
 \centering
\small{
\begin{tabular}{l c c c}
\hline
Compiler & \textbf{GFortran} & \textbf{PGIF90} & \textbf{XLF}\\
Version  & 8.1.0 & 18.10 & 16.1.1.2 \\
\hline
\end{tabular}
}
 \label{tab:compilers}
\end{table}

In all our experiments in this section, we have used OpenMPI 3.0.0 and different compilers, which Table~\ref{tab:compilers} summarizes.

The results shown are the average of 20 time steps, removing the first one of the simulation because it includes some initialization overheads. 
We present different measurements of the duration of the whole time step and also split by the most time-consuming phases. 
As we explained in Section \ref{sec:numer}, and illustrated by the trace of Figure \ref{fig:trace_phases}, there are three phases: 
the element assembly, the boundary assembly, and the algebraic solver. 
In the present test case, the most computation-intensive phase is the element assembly, representing 75\% of the total computation, as shown in Table~\ref{tab:counters}.

We divide the analysis of CPUs into three parts: optimum pack size for CPUs, compiler comparison and performance impact of threading.

\subsection{Optimum pack size}
\label{sec:opt_pack_cpu}

As we have explained in the previous section, to improve the performance, the elements are stored and computed in packs. 
The size of these packs can have a significant impact on performance. 
In this section, we are going to analyze this impact and find the optimum value to use it in the following experiments.

We have performed the experiments for this subsection on 10 nodes of the \power{}, launching 40 MPI processes per node (i.e., one MPI process per core and not using SMT).
We have compiled Alya using GFortran 8.1.0, and the test case is the airplane simulation of 31.5M elements.

\begin{figure}[h!tbp]
  \centering
  \includegraphics[width=0.65\textwidth]{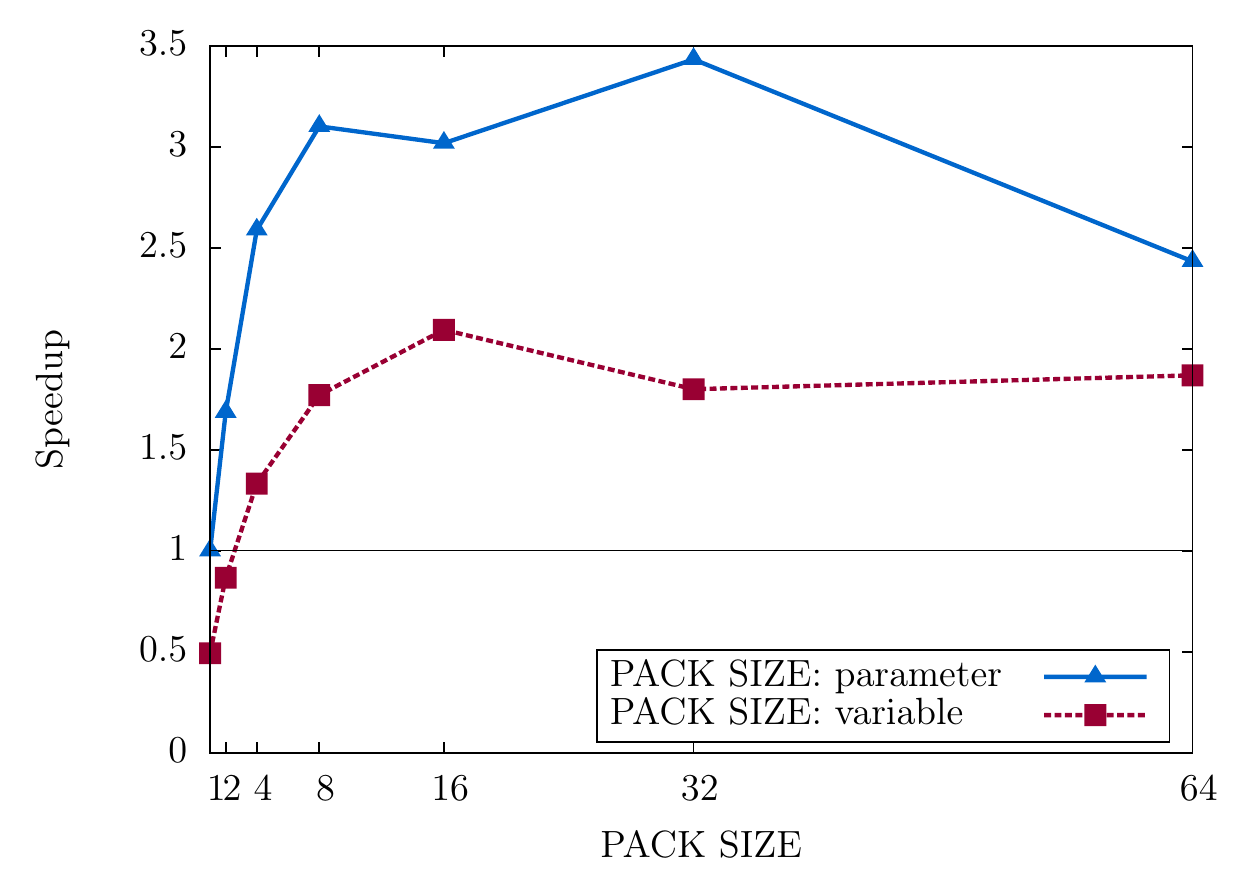}
  \caption{Speed up for different pack sizes, for {\tt PACK\_SIZE} defined as a Fortran parameter or a variable.}
  \label{fig:pack_size} 
\end{figure}

Figure~\ref{fig:pack_size} shows the speedup obtained when using different
values for the pack size. The speedup is computed according to the execution time using a pack size
 of 1 as a compilation parameter. The X-axis represents the different sizes for the pack. 
If we observe the line corresponding to the pack size defined as a variable, we are evaluating the
 improvement in performance due to the locality of the data, and it will be related to the length
 of the cache line for the last level cache. 
The optimum pack size, when used as a variable, is 16 in this architecture, for
larger pack sizes the performance degrades since are too far from the POWER9
vector length.

If we look at the pack size defined as a compilation parameter, we are evaluating the combined benefit of the better data locality and the better use of the vector units. 
Defining the value at compile provides more information to the compiler that
becomes more efficient in the vectorization of the corresponding loop.
For this reason, the performance of the pack size defined as a compilation parameter is always better than the same size defined as a variable. 
In the case of the pack size defined as a parameter, the size that delivers a higher performance is 32. 
It is interesting to see that the optimum value when using a parameter does not match the optimum value when defining the pack size as a variable.
It means that even a better configuration would be possible if the hardware structure in terms of cache line and vector length would match to exploit both benefits.

\subsection{Compiler comparison}

To compare the performance of the code using different compilers we have simulated the airplane with the coarse
mesh of 31.5M elements, using 12 nodes of the \power{}. Equivalent optimization
flags have been used in the compiling process. 
In the experiments, we have used four configurations of SMT per core launching 1, 2, 3 or 4 MPI processes per core. 
We have measured the elapsed time for the three main kernels of the code, namely the element
assembly (the most time-consuming kernel), the boundary assembly and algebraic solver. In addition,
we have measured the elapsed time of a complete time
step, which gives an idea of the overall performance of the code. 
In all the cases we report time in seconds.

\begin{figure}[h!tbp]
  \centering
  \includegraphics[width=0.49\textwidth]{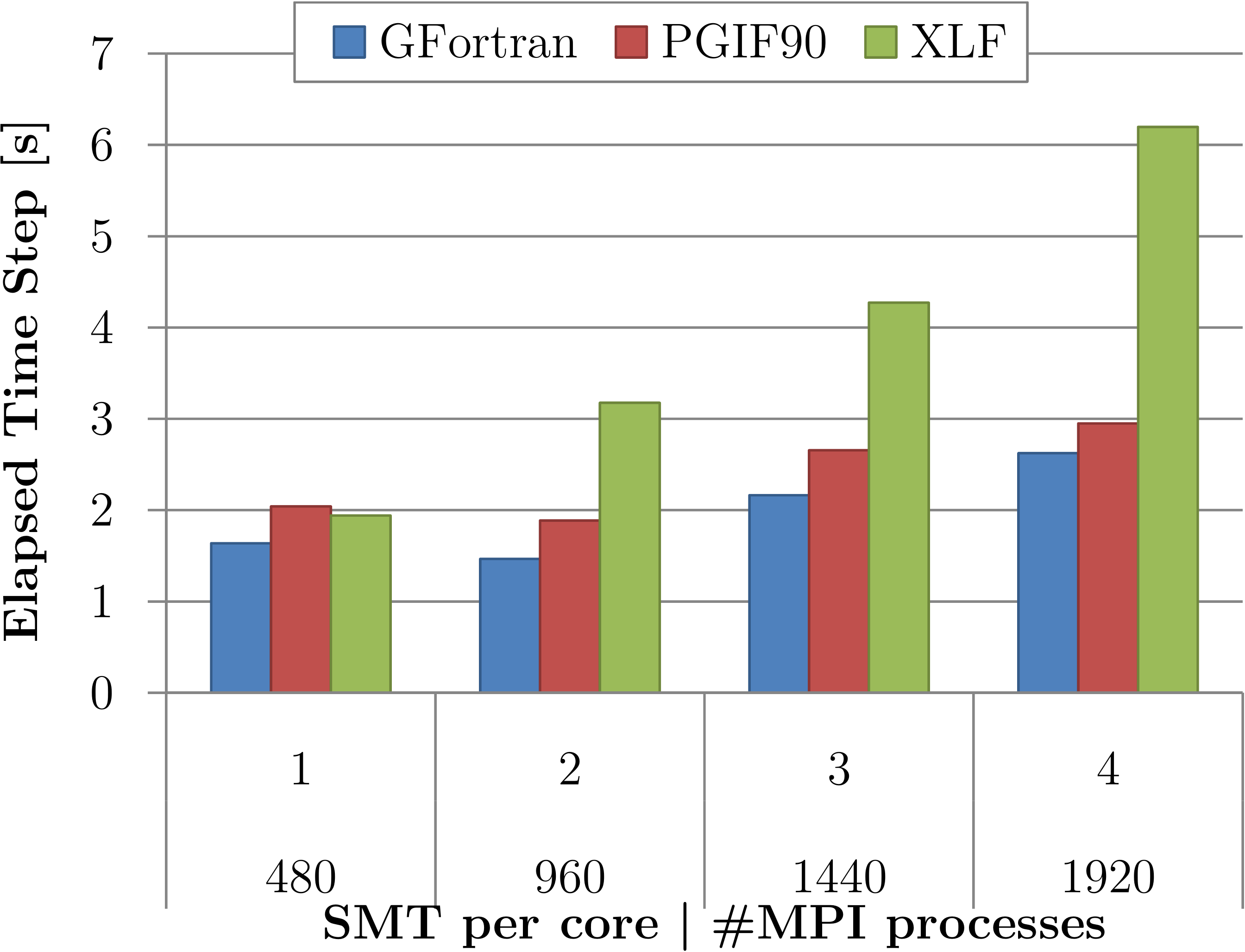}
  \includegraphics[width=0.49\textwidth]{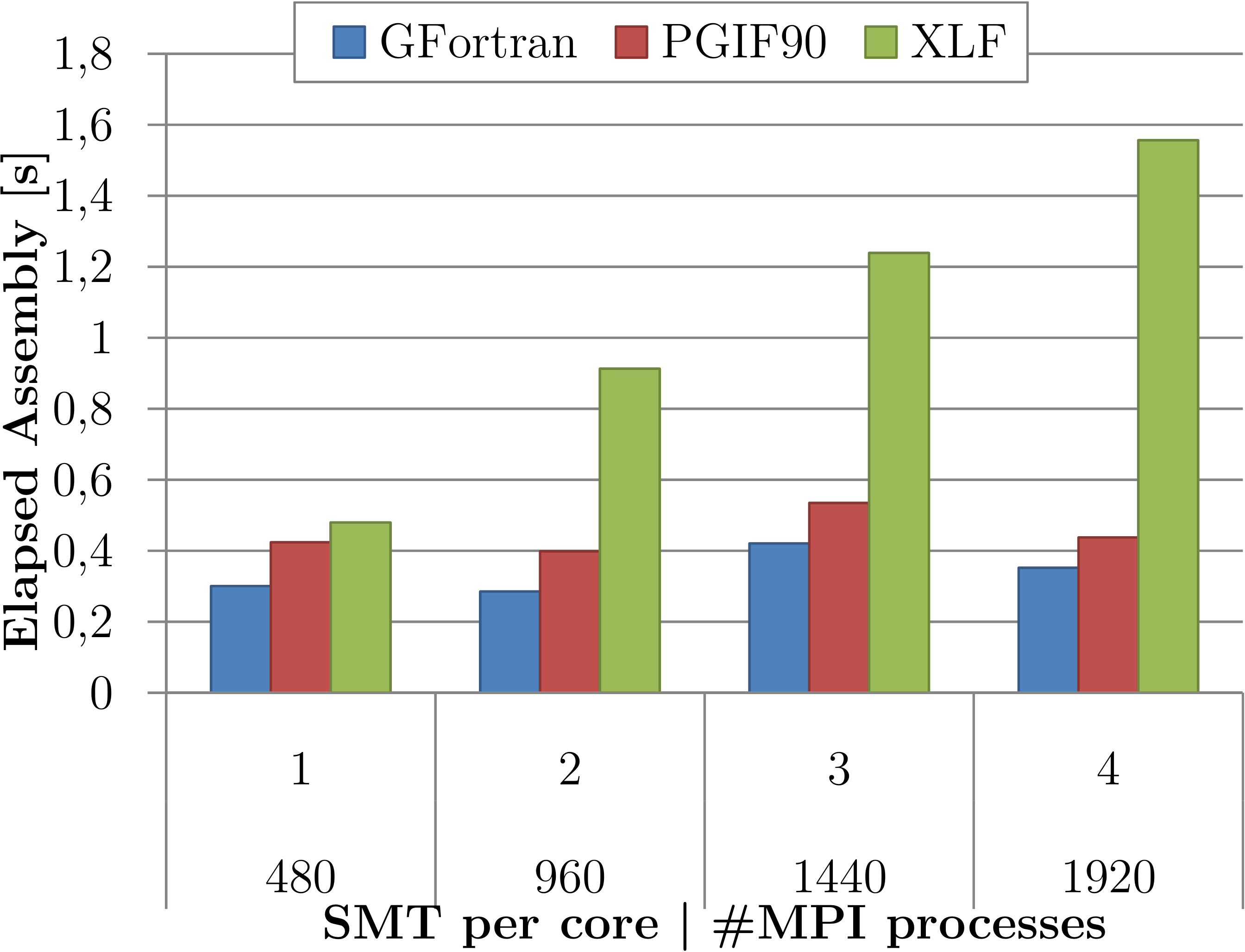}
  \caption{Compiler comparison for elapsed time. Left: one complete time step. Right: element assembly.}
  \label{fig:compiler} 
\end{figure}

Figure~\ref{fig:compiler} shows the elapsed time per time step at the left-hand side and the elapsed time in the element assembly phase on the right-hand side. 
The X-axis represents the number of SMT used per core and the total number of MPI processes launched for the simulation. 
We can observe that both charts have a similar trend because the element assembly phase is the kernel that has the highest weight in terms of computational load. 

We conclude that GFortran is the compiler that obtains the best performance in all the configurations. 
We can also note that PGIF90 performs slightly worse than GFortran, but the trend in the different configurations is almost identical to the GFortran one. 
On the other hand, XLF obtains a similar performance to PGIF90 and GFortran when launching one MPI process per core (i.e., using one SMT per core)
but it degrades fast when increasing the number of SMT used per core. 

Regarding the performance of using several SMT per core in the element assembly, 
we conclude that with GFortran and PGIF90 using two SMT per core gives a slightly better performance than using one SMT, 
but when increasing to 3 and 4, there is no benefit in increasing the number of SMT.

\begin{figure}[h!tbp]
  \centering
  \includegraphics[width=0.49\textwidth]{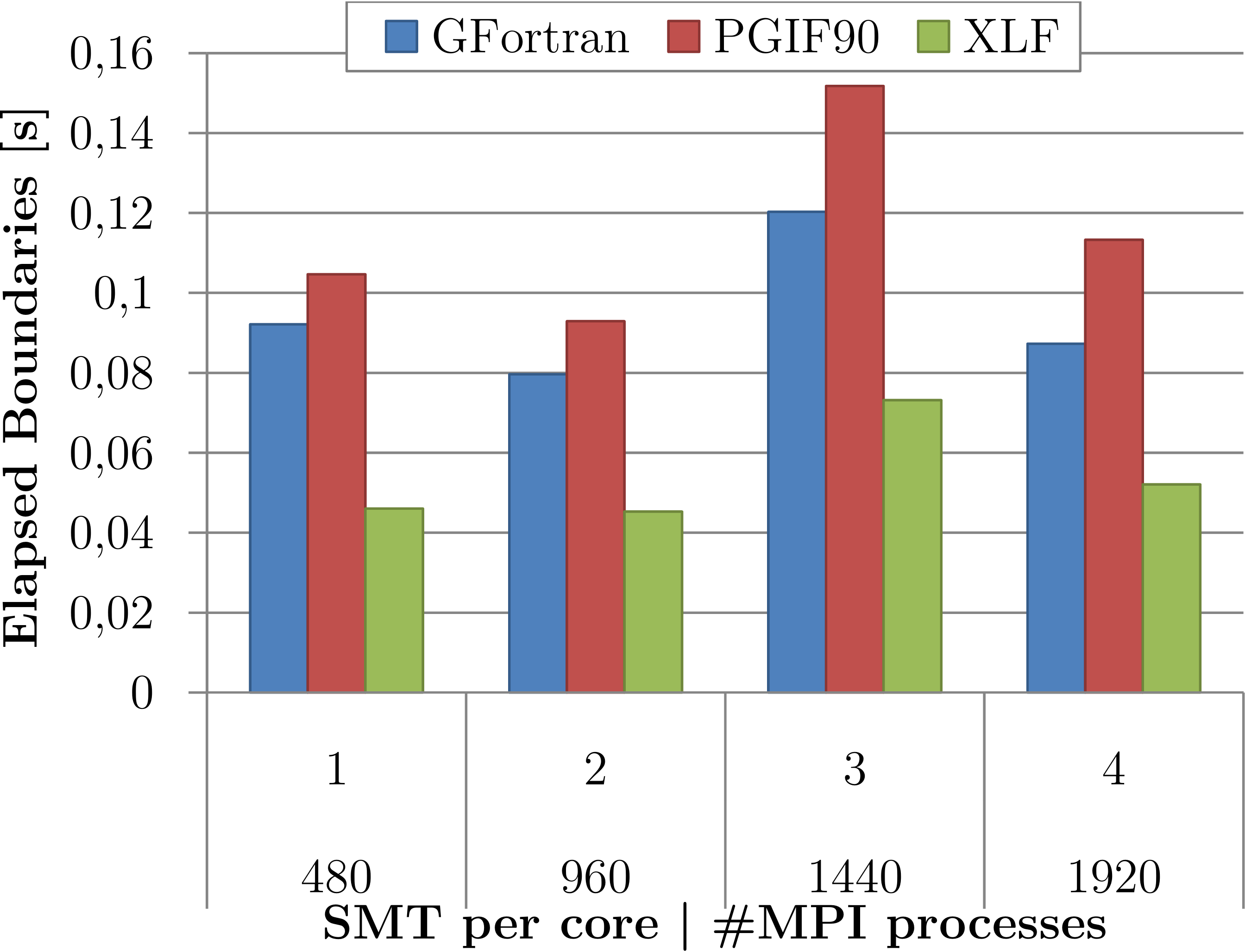}
  \includegraphics[width=0.49\textwidth]{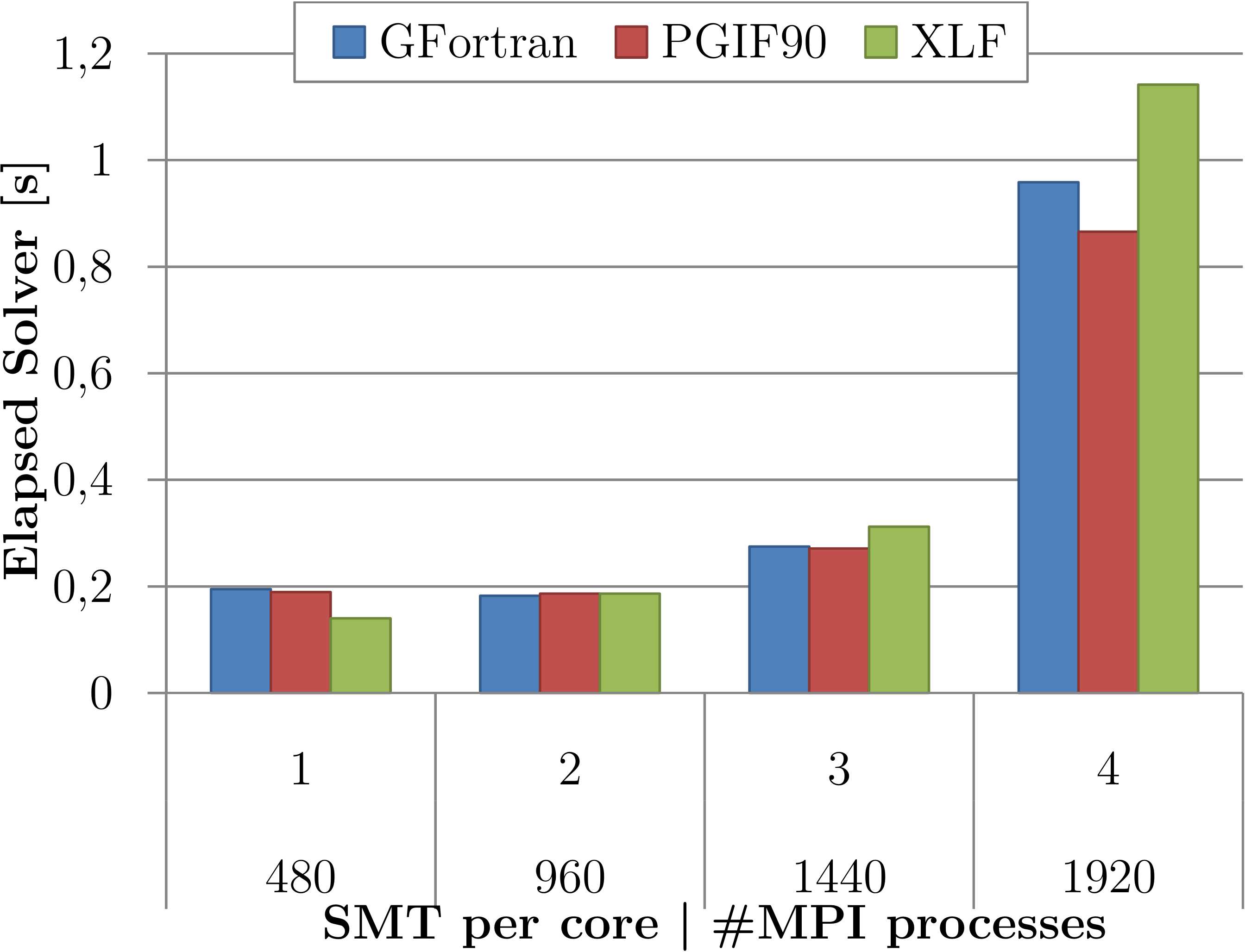}
  \caption{Compiler comparison for elapsed time. Left: boundary assembly. Right: algebraic solver.}
  \label{fig:compiler2} 
\end{figure}

In the left-hand side of Figure~\ref{fig:compiler2}, we can see the compiler comparison for the boundary assembly. 
In this kernel, the best performance is provided by the XLF compiler in all the numbers of SMT. 
Also, GFortran performs slightly better than PGIF90 in all the cases, but again, the trend of their performance is very similar. 
For the boundary assembly, the best configuration is to use 2 SMT per core independently of the compiler used, and the worse case is to use 3 SMT per core.

The right-hand side of Figure~\ref{fig:compiler2} shows the compiler comparison for the algebraic solver. 
In this case, the best option is to use XLF with 1 SMT per core. For GFortran
and PGIF90 it is better to use 2 SMT per core. 
When using 2 or 3 SMT per core the performance of the three compilers is very similar. 
When using 4 SMT all of them degrade the performance, being almost 4 times slower then using 1 SMT.

From this analysis we can see that XLF is generating the code that delivers the worse performance for the element assembly and the best one for the boundary assembly. 
To investigate the performance loss in the element assembly with XLF, we obtained hardware counters from the different executions, and we
observe that the executions with XLF present a worse Instructions per Cycle
(IPC) ratio and a significant decrease in the number of cycles per microsecond used by each process when increasing the number of SMT per core. 
We need to analyze further this effect as it seems that the processor is decreasing the frequency during the execution of that part of the code.

\subsection{Threading performance}

In this subsection, we study the impact on the performance of using different threading levels; SMT and OpenMP.

In all the experiments of this section, we simulate the airplane of 31.5M elements on 12 nodes of the \power{}. 
Note that we have fixed the number of resources for all the tests. What changes is the distribution of MPI processes and OpenMP threads combined with the number of SMT threads enabled per core. 
All the experiments of this study use GFortran 8.1.0 and OmpSs as a parallel runtime for OpenMP.

\begin{figure}[h!tbp]
  \centering
  \includegraphics[width=0.49\textwidth]{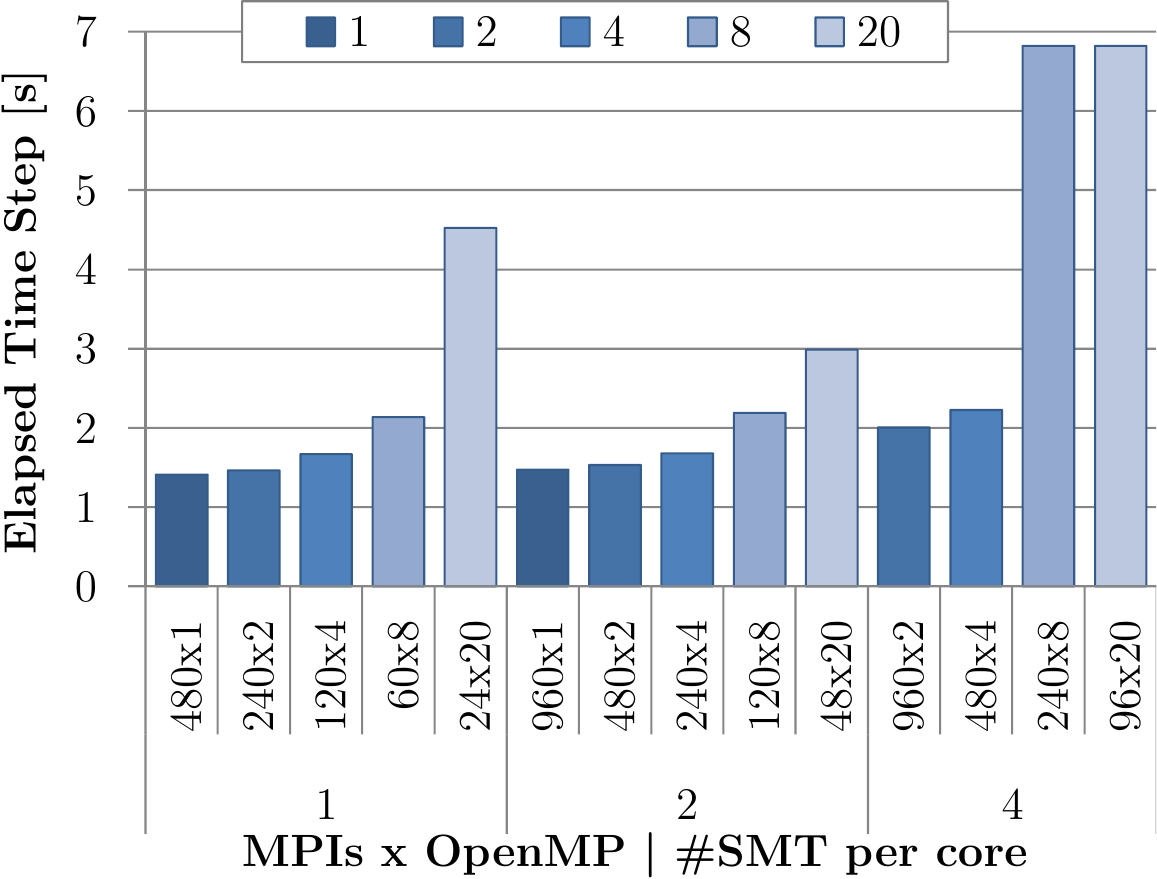}
  \includegraphics[width=0.49\textwidth]{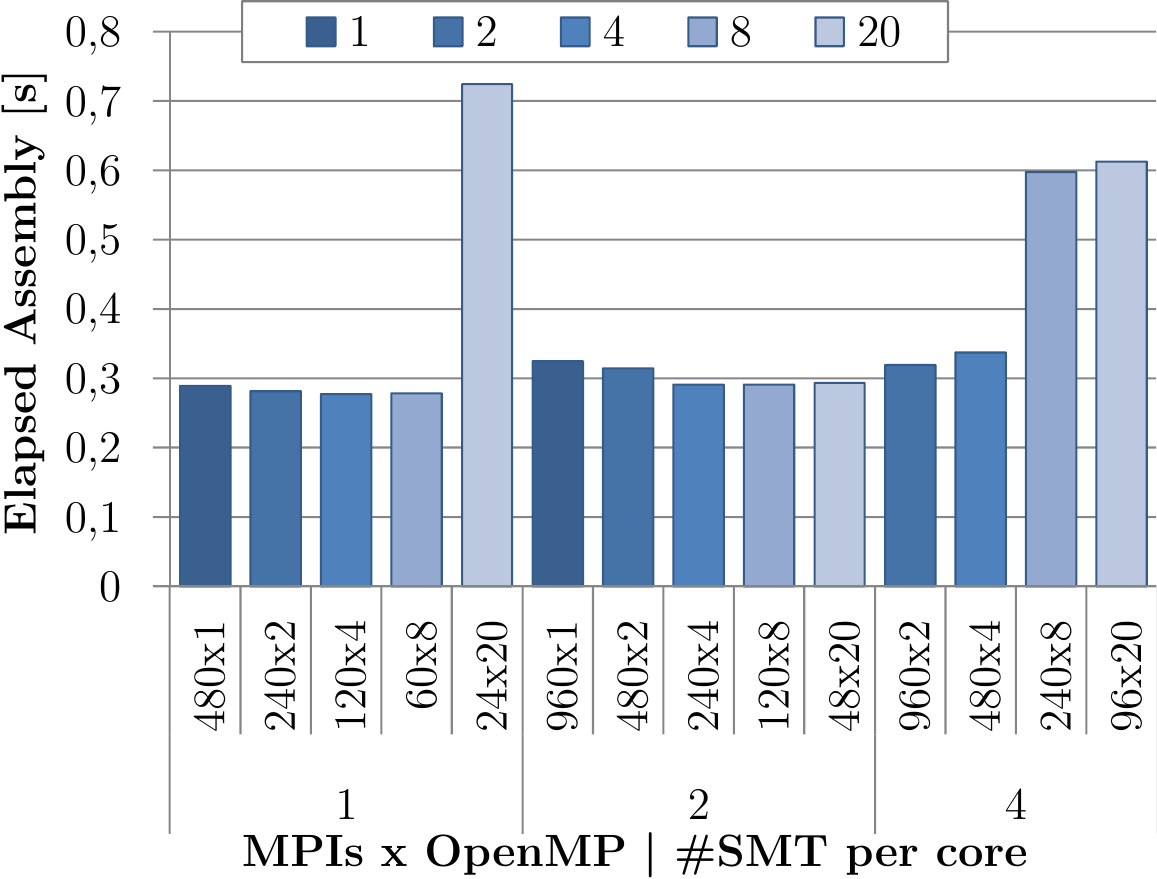}
  \caption{Threading comparison for elapsed time. Left: one complete time step. Right: element assembly.}
  \label{fig:threading} 
\end{figure}

Figure~\ref{fig:threading} shows the elapsed time in seconds for the different executions on 12 nodes. 
The X-axis represents the hybrid configuration in the form of: 
``number of MPI processes'' x ``number of OpenMP threads per process'', the second level of the axis indicates the number of SMT per core. 
For the sake of clarity, the colors of the bars correspond to the numbers of OpenMP threads used.

The right side of the figure shows the results obtained for the element assembly. 
We observe that the best configuration consists in using only one SMT per core. 
In this case, the use of OpenMP threads scales up to 8 OpenMP threads per MPI process. 
When using 2 SMT threads per core, the performance is close to using 1 SMT per core, and the use of OpenMP threads also scales up to 20 OpenMP threads per MPI process. 
But when enabling 4 SMT per core the trend is different, increasing the number of threads degrades the performance and OpenMP only scales up to 4 threads per MPI process.

\begin{figure}[h!tbp]
  \centering
  \includegraphics[width=0.49\textwidth]{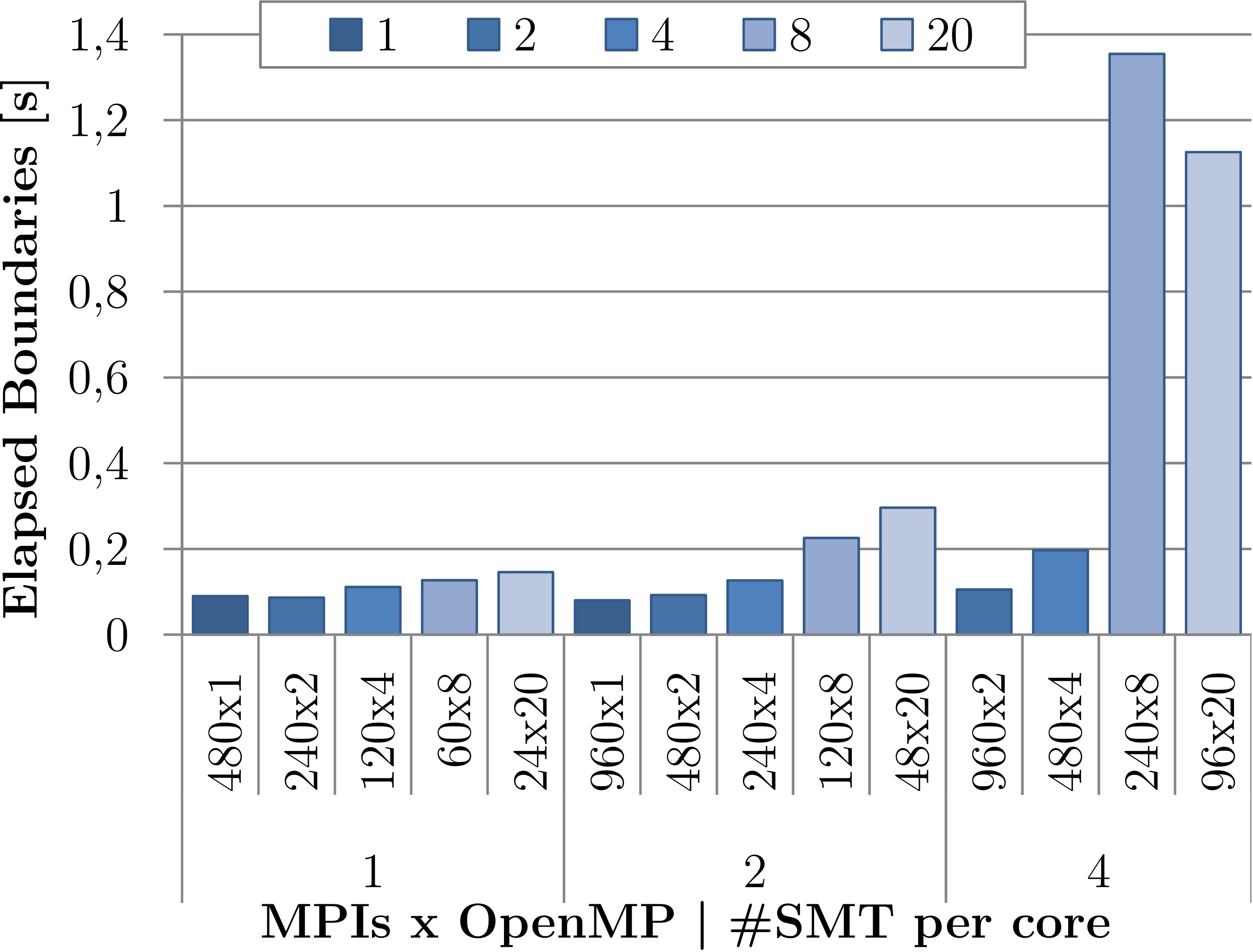}
  \includegraphics[width=0.49\textwidth]{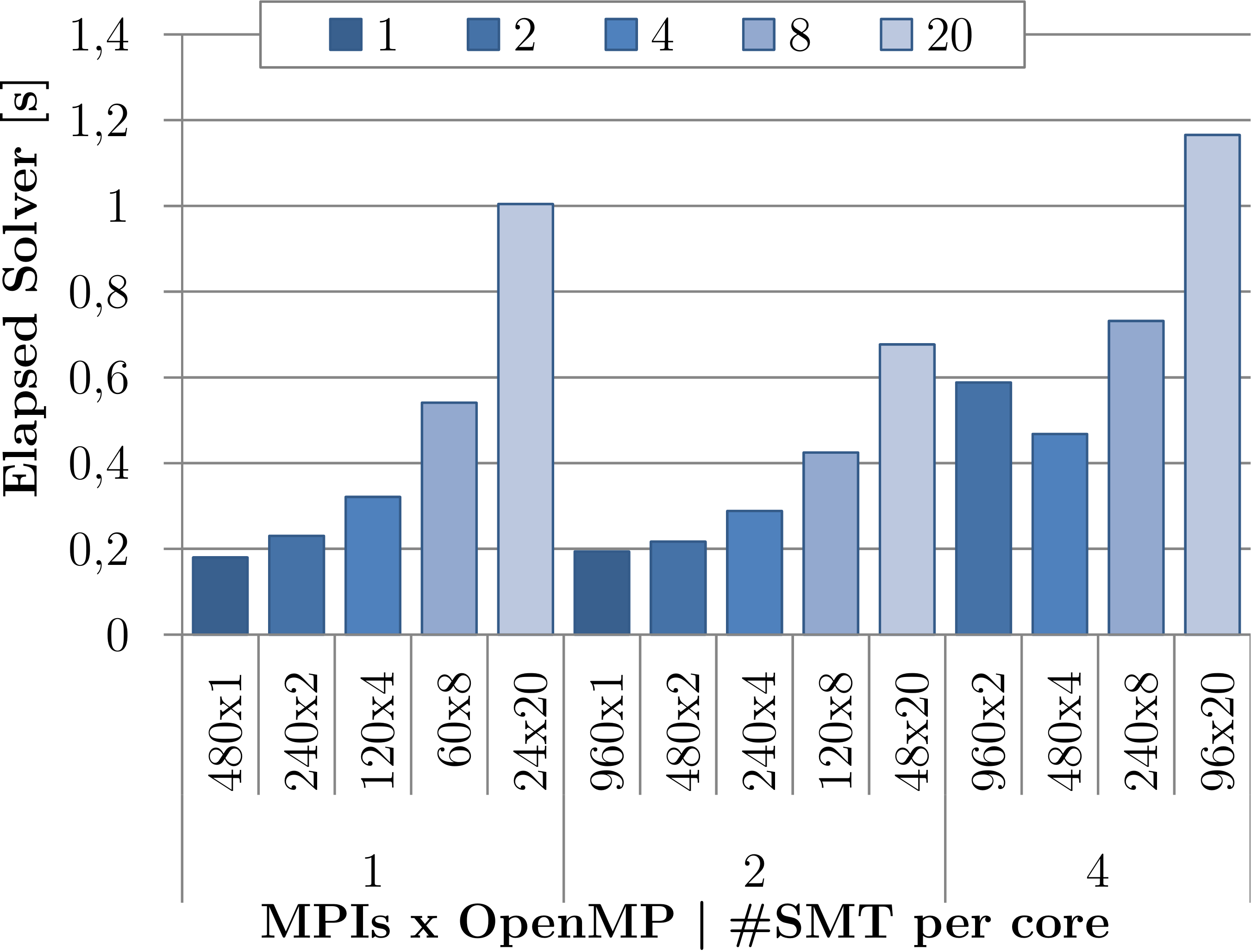}
  \caption{Threading comparison for elapsed time. Left: boundary assembly. Right: algebraic solver.}
  \label{fig:threading2} 
\end{figure}

Figure~\ref{fig:threading2} shows the elapsed time for the boundary assembly (left) and the algebraic solver (right) when using different levels of threading. 
For the boundary assembly computation, the best configuration is to spawn one OpenMP thread per MPI process and use 2 SMT per core. 
The trend and performance when using 1 or 2 SMT per core are similar. 
The best option is to use 1 or 2 OpenMP threads per MPI process. 
When using 4 SMT per core the performance with 2 OpenMP threads is close to the one obtained using 1 or 2 SMT per core.

The right-hand side of Figure~\ref{fig:threading2} shows the elapsed time of the solver. 
The best option to reduce the execution time of the algebraic solver is to run with 1 OpenMP thread per MPI process and use one SMT per core. 
In the case of the solver, we observe that using 1 or 2 SMT per core gives similar performance, but when using 4 SMT per core the performance drops. 
The parallelization with OpenMP does not scale for any of the levels of SMT; the execution time grows with the number of OpenMP threads used per MPI process.

We can conclude that finding an optimum configuration for a complex application is not a trivial task.
We have seen that the different kernels obtain different performances for the same feature or configuration.
Moreover, for a specific input, one option is to consider at the global execution time of a complete time step and find the best configuration, but we cannot suggest a configuration that will obtain the best performance for any simulation.
Because the weight of each kernel is not constant for all the input cases, it
depends highly on the physics solved and the parameters set to perform the
simulations.

\subsection{Scalability}

To finish this section, we present some scalability results using the CPUs. 
For these experiments, we have used the 31.5M mesh of the airplane, and we always use full nodes. 
We have used 3 different configurations: the two best options from Figure~\ref{fig:threading} left, 
being pure MPI with GFortran and the hybrid version with GFortran and OmpSs with 2 threads always with 1 SMT per core, 
and the best configuration from Figure~\ref{fig:threading} right, using the hybrid version with GFortran, 8 OmpSs threads and 1 SMT per core. 
We test the best configuration for the whole time step and the element assembly.

\begin{figure}[h!tbp]
  \centering
  \includegraphics[width=0.49\textwidth]{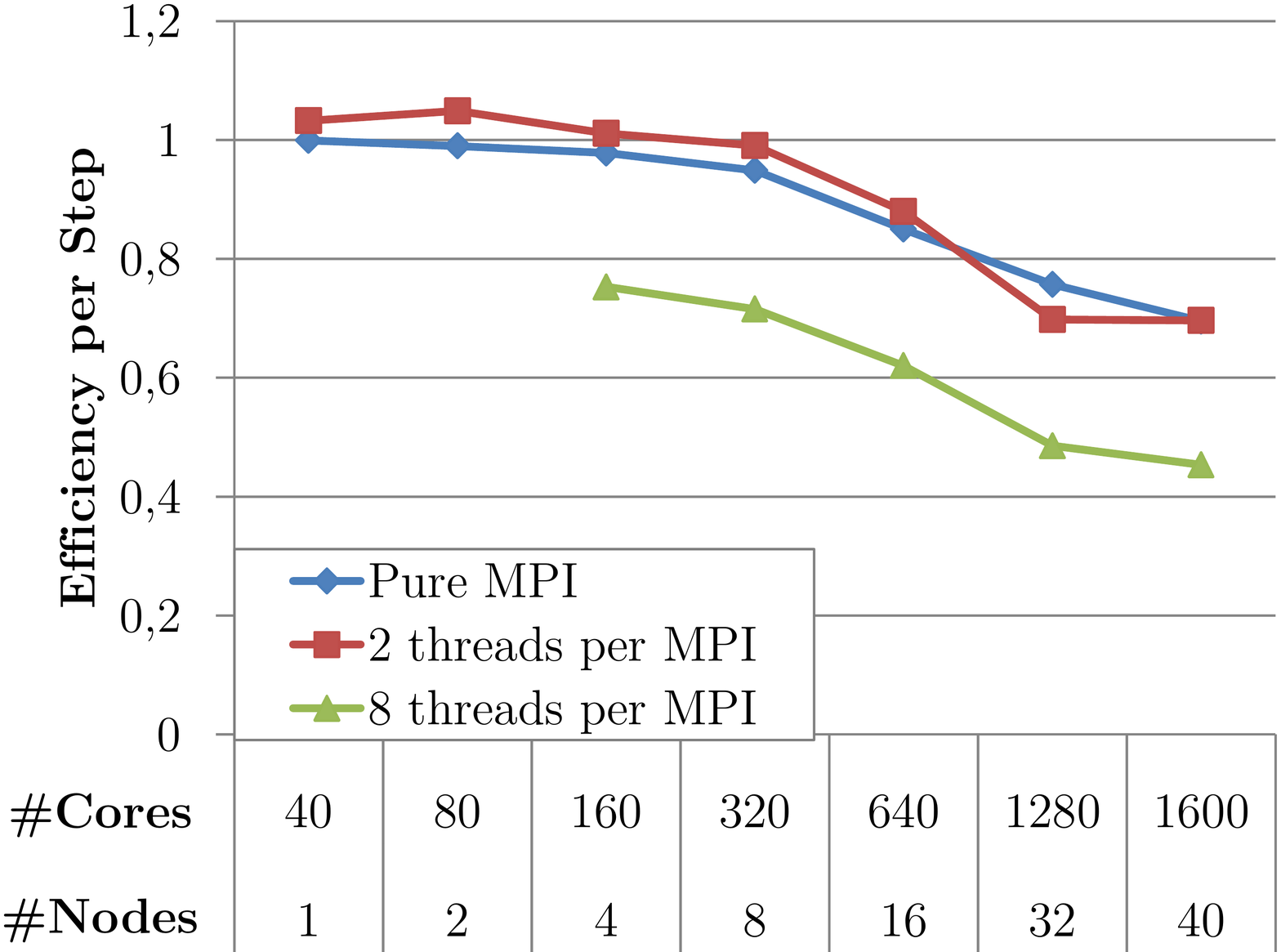}
  \includegraphics[width=0.49\textwidth]{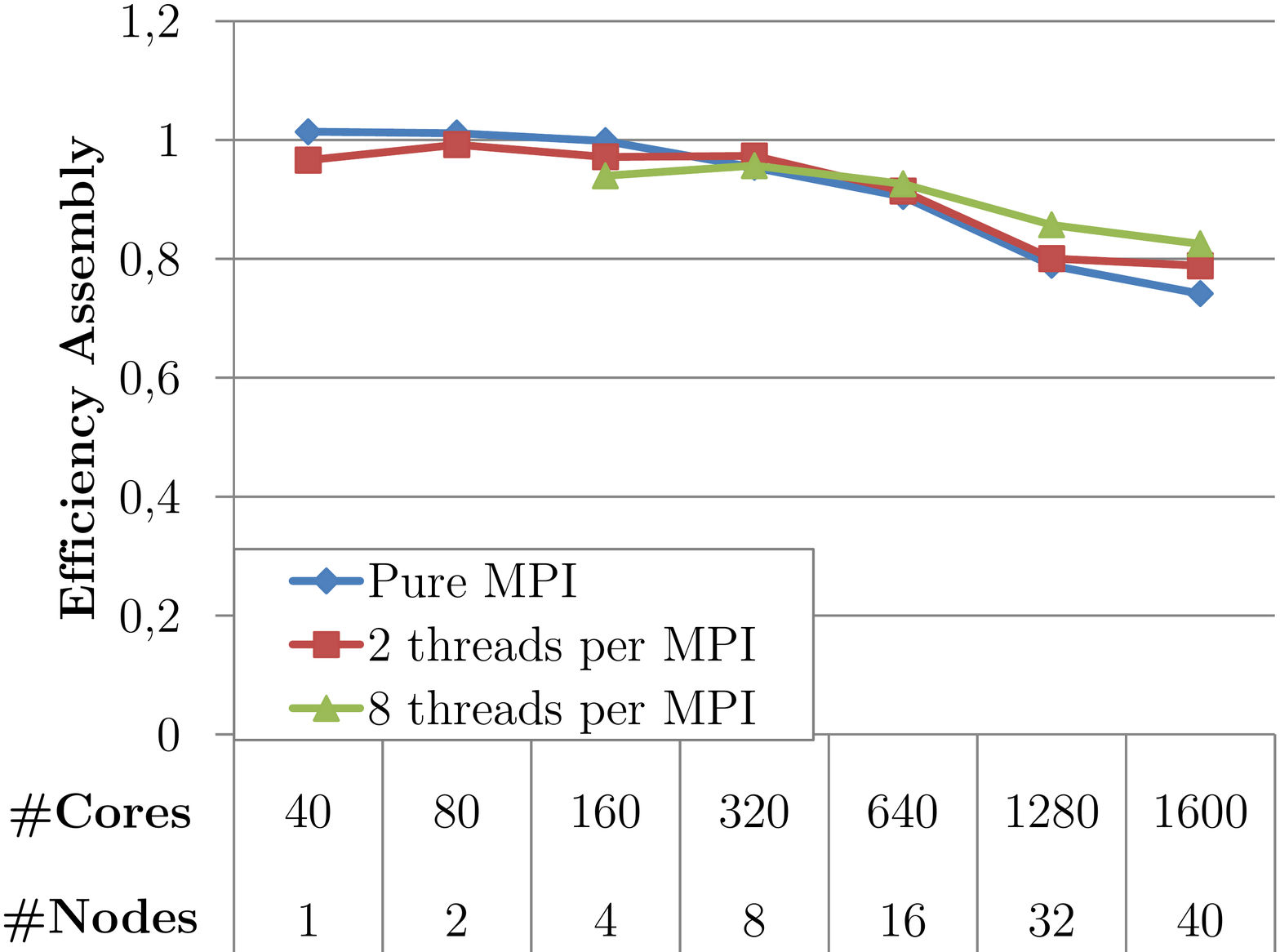}
  \caption{Efficiency measuring the elapsed time. Left: one complete time step. Right: element assembly.}
  \label{fig:Efficiency1} 
\end{figure}

In Figure \ref{fig:Efficiency1} we can see the scalability obtained up to 40
nodes of the \power{}. We show the Efficiency computed as:
$Eff_x=\frac{t_1}{t_x/x}$, being $t_x$ the elapsed time using $x$ nodes, as reference ($t_1$) we have used the elapsed time in one node with the pure MPI version. All the numbers have been obtained averaging twenty time steps of two different runs

The left-hand side of Figure~\ref{fig:Efficiency1} shows the efficiency obtained for the whole time step.  
We can observe that the version with 2 OmpSs threads obtains the best efficiency up to 16 cores. 
The efficiency of the pure MPI version is above $80\%$ when using up to 16 nodes also. 
On the other hand, the performance of the 8 threads version is below $80\%$ in all the cases. 

The right-hand side of Figure~\ref{fig:Efficiency1} shows the efficiency of the
assembly phase. We observe that the efficiency obtained by this kernel is above
all the other kernels, being in almost all the cases above $80\%$. As we explained in Section~\ref{sec:perf_char} the element assembly phase does not include any internal communication. For this reason, all the efficiency loss at the MPI level is due to the load imbalance. In Section~\ref{sec:loadbalancing}
we will address this issue further and present the balancing mechanism implemented to solve it.

\begin{figure}[h!tbp]
  \centering
  \includegraphics[width=0.49\textwidth]{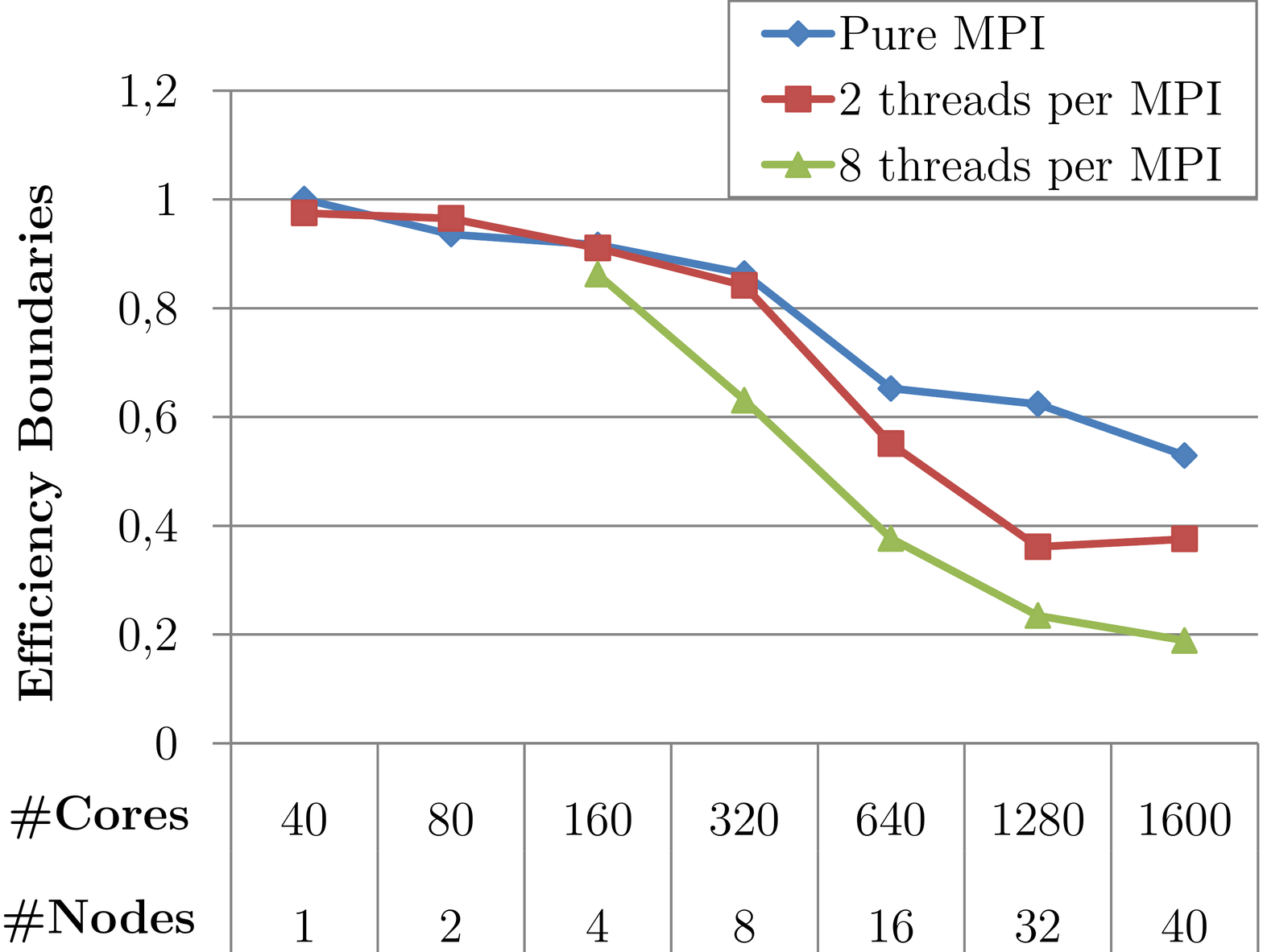}
  \includegraphics[width=0.49\textwidth]{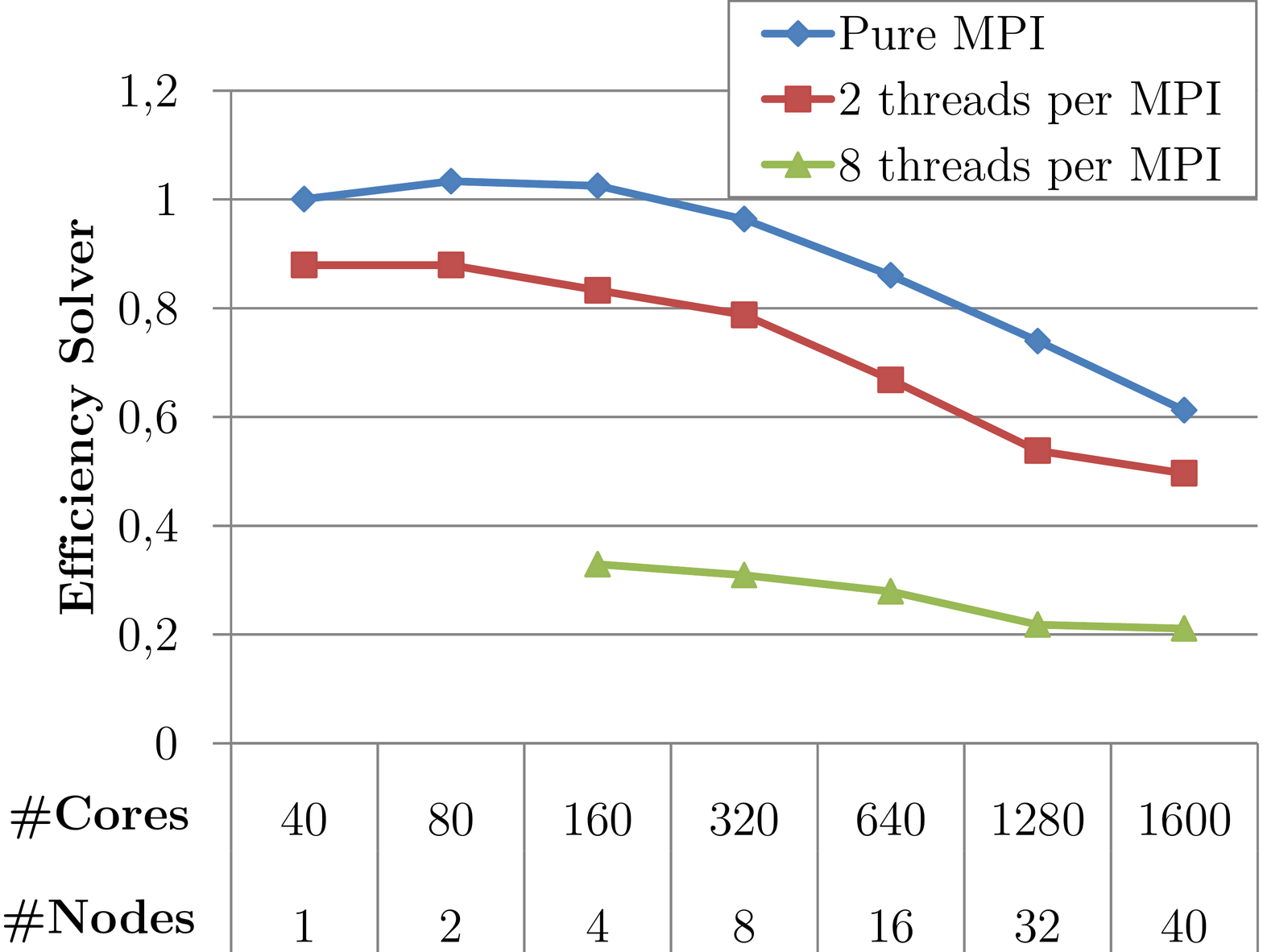}
  \caption{Efficiency measuring the elapsed time. Left: boundary assembly. Right: algebraic solver.}
  \label{fig:Efficiency2} 
\end{figure}

The left-hand side of Figure~\ref{fig:Efficiency2} shows the efficiency obtained by the assembly of boundary assembly phase. 
We can observe that up to 8 nodes the efficiency is above $80\%$ when using 2 OmpSs threads or pure MPI. 
When using more than 8 nodes, the efficiency drops reaching almost $50\%$ in the
case of the pure MPI and $40\%$ in the case of 2 threads per MPI process.

The right-hand side of Figure~\ref{fig:Efficiency2} shows the efficiency of the algebraic solver. 
We can see that the efficiency of this kernel is good with the pure MPI version
up to 16 nodes, showing an efficiency above $80\%$. 
As we have observed before the use of OmpSs threads penalizes the performance of the algebraic solver.

%% file: sections/05_gpu_performance.tex
\section{GPU performance analysis}
\label{sec:gpu}

This section is devoted to the performance analysis of the GPU implementation of Alya
 considering the different optimizations explained in Section~\ref{simdsimt}. 
We obtained the numerical results by evaluating the performance in one node of the \power{},
i.e., using the four NVIDIA V100 GPU results and comparing it with the two IBM
POWER9 processors of the node. 

\begin{figure}
\centering
  \includegraphics[width=.45\textwidth]{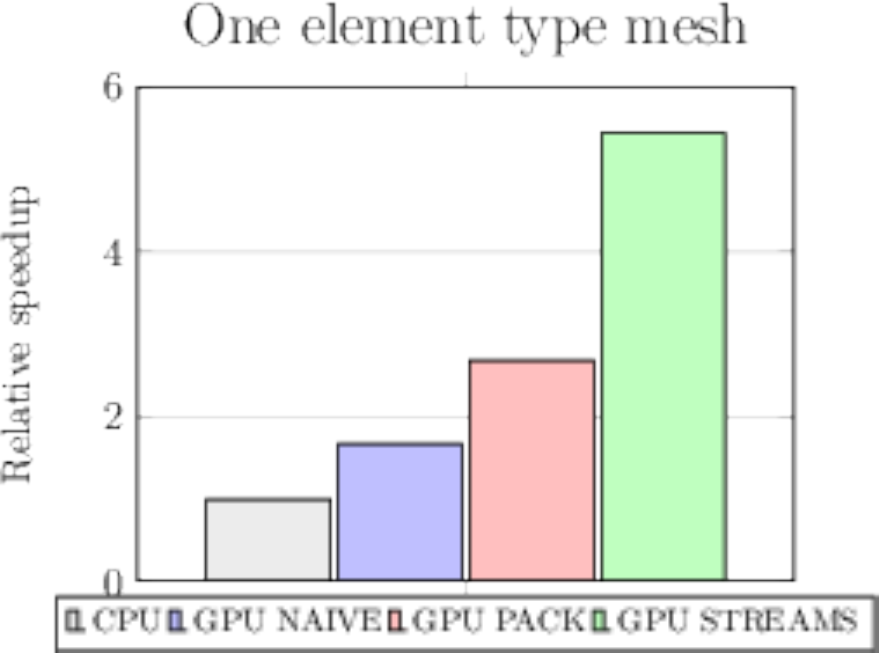} 
  \includegraphics[width=.45\textwidth]{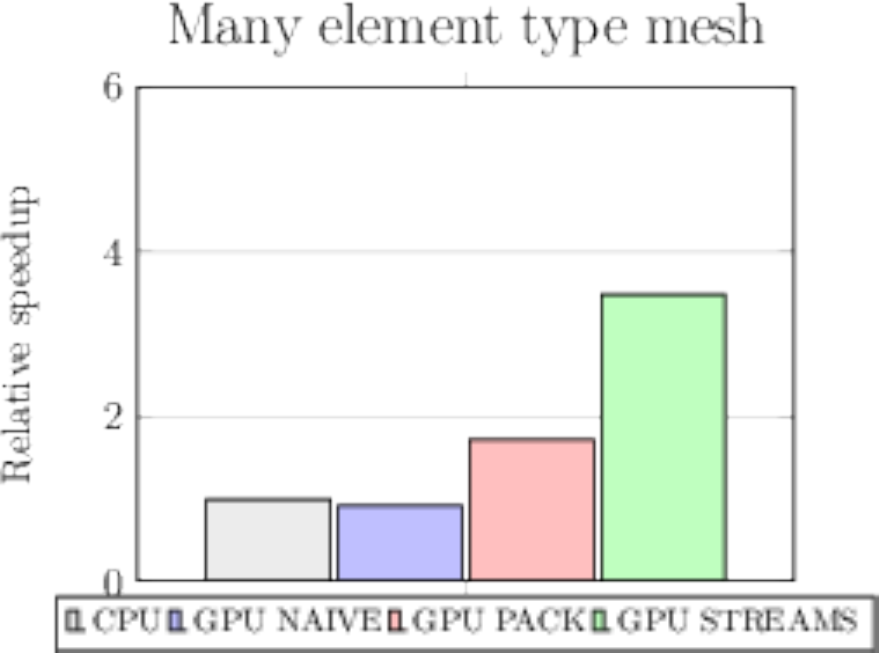}
   \caption{Assembly speedup of 4 GPUs with respect to the 40 cores.
Left: tetrahedral mesh  sphere simulation. Right: many kind of element mesh airplane simulation.}
   \label{fig:gpuassembly_perf}
 \end{figure}

\begin{figure}
\centering
  \includegraphics[width=.35\textwidth]{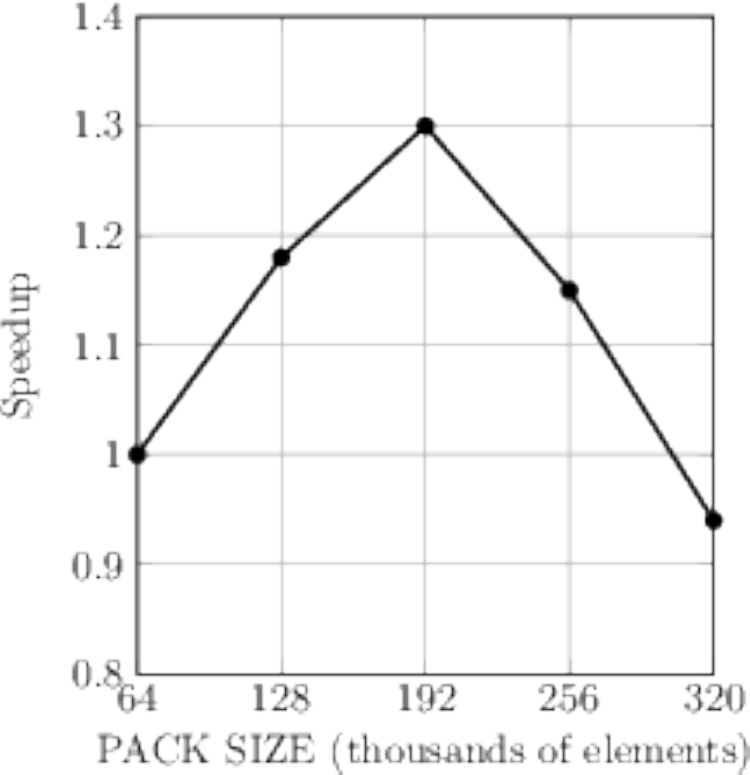}
  \includegraphics[width=.45\textwidth]{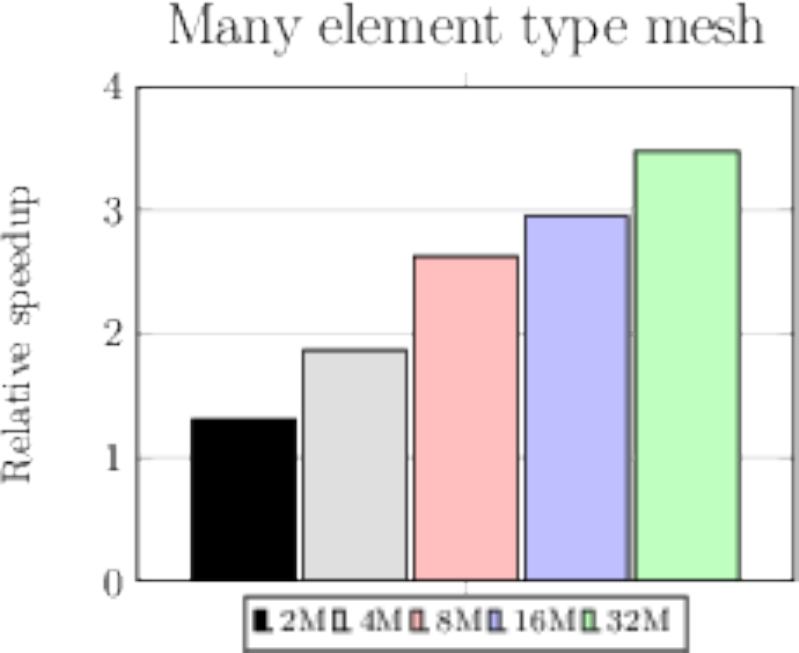}
   \caption{Left: Speedup for the different pack sizes. Right: Speedup of 4 GPUs with respect to the 40 cores in the airplane simulation for different
     workloads.}
   \label{fig:gpuassembly_spd}
 \end{figure}

Figure~\ref{fig:gpuassembly_perf} depicts the performance improvement of the GPU 
implementation by gradually introducing the optimizations. 
We used the best CPU version of the code as the baseline for comparing the GPU performance. 
Also, we have included a simulation of the flow around a sphere (only tetrahedral
elements mesh) to quantify the impact of 
using meshes with different type of elements such as the one in the airplane
 simulation (tetrahedron, prism, hexahedron, etc.).
Note that additional data structures are needed to deal with the extra
complexity derived from the many kind of elements in the mesh. 
The first GPU execution is based in a straightforward OpenACC implementation using the same
 configuration than the CPU. 
Such version outperforms the CPU implementation when using the sphere simulation ($1.67\times$), 
however, for the airplane simulation it decreases performance due to low
occupancy of the GPUs.
The third column in the plots corresponds to finding the optimal parameters {\tt PACK\_SIZE}
 and \texttt{vector\_length} for the GPU.
The selection of the optimal parameters is shown in
Figure~\ref{fig:gpuassembly_spd} (left). The optimal number of threads is obtained using the CUDA Occupancy calculator.
In the airplane simulation the optimal {\tt PACK\_SIZE} is 196,608  and the \texttt{vector\_length} is 128. 
Such configuration means that the GPUs fetches calculations in groups of kernels of 1536 blocks
 and 128 threads. The relative speedup obtained by tuning these parameters is of $2.68$ and $1.72$ times
 for the tetrahedral and the many kind of elements meshes respectively. 
Finally, we represent the pipelined execution model in the last column of the plots. 
Finding the optimal number of streams per each GPU is limited by the memory footprint
 and access pattern of the kernel. 
The optimal airplane simulation kernel uses a large {\tt PACK\_SIZE}, and therefore
 it has large memory requirements. 
As a result, the optimal number of streams per GPU is four, and the pipelined implementation
 outperforms all previous trials. 
The final speedup of the GPUs according to the CPUs of the node is of $5.44$ and $3.48$ times
 for the sphere and airplane meshes respectively. 
These results match other unstructured CFD codes running on
GPUs~\cite{OYA18TAYLOR,OYA13}.

Figure~\ref{fig:gpuassembly_spd} (right) shows the speedup of the best GPU implementation according to the
 optimal CPU version at the node level. 
The bars in the plot represent different workloads that range from 2 million up to
 32 million elements per node.
Note that increasing the workload benefits the GPU performance since the devices need
 a certain level of occupancy to reach its optimal execution.
The acceleration obtained by the GPUs ranges from 1.3 up to 3.48 times faster than the CPUs. 

%% file: sections/06_coexecution.tex
\section{Mesh partitioning and load balancing}
\label{partition}

\subsection{Mesh partitioning}
\label{sec:meshpartitioning}

Mesh partitioning is traditionally formulated as a graph partitioning problem, that is a well-studied NP-complete problem generally addressed using multilevel heuristics. Publicly available libraries such as ParMetis~\cite{Karypis:98} or PT-Scotch~\cite{CHEVALIER2008318} 
implement different variants of them. 
Geometric partitioning techniques are another alternative, available for instance in the Zoltan library~\cite{Zoltan2012}. They obviate topological interactions between mesh elements and perform the partition considering only their spatial location. 
A SFC is generally used to map the mesh elements into a 1D space, which is then split it into equally weighted sub-segments. 
There are different definitions of SFC, such as the Hilbert or Peano curves \cite{Sagan1994}. 
The common goal is to preserve locality, i.e., preserve the proximity between points in their 1D projections and vice versa. 
A significant advantage of the geometric approach is that it is highly scalable~\cite{LIEBER2018575}. 
However, while load balancing can be easily achieved, the interface between subdomains, 
measured in terms of edge-cuts in the graph methods, is not explicitly minimized
but it is only treated indirectly through the SFC locality property. 

In this paper, we use an in-house SFC-based mesh partitioner that we already described in detail in our previous work~\cite{BORRELL2018}. 
Table~\ref{argsfc} presents the inputs of the algorithm.
Regarding the third argument, the location of the mesh elements $C[...]$, for each element we provide the average of the coordinates of its nodes. 
On the other hand, the weights $W[...]$, provided as the fourth argument, are used to deal with elements of different type and different associated computing cost. 
To estimate this cost, we use a simple model: the weight associated with an element is equal to its number of Gauss points.
However, in many situations, this model may be imprecise to obtain well-balanced distributions. 
In section~\ref{sec:loadbalancing}, we present a novel method to address this problem, based on partition correction coefficients.

\begin{table}
 \caption{Input arguments of the SFC-based in-house partitioning algorithm.}
\small{
\begin{tabular}{ l l } 
 \hline
 $N$       & Number of mesh elements  \\ 
 $P$       & Number of desired partition subsets  \\ 
 $C[...]$  & Location of each mesh element; array of dimension $3 \times N$  \\ 
 $W[...]$  & Weight associated to each mesh element; optional array of dimension $N$  \\
 \hline
\end{tabular}
}
\label{argsfc}
\end{table}	

\begin{figure}[!htb]
  \centering
    \includegraphics[width=0.5\textwidth]{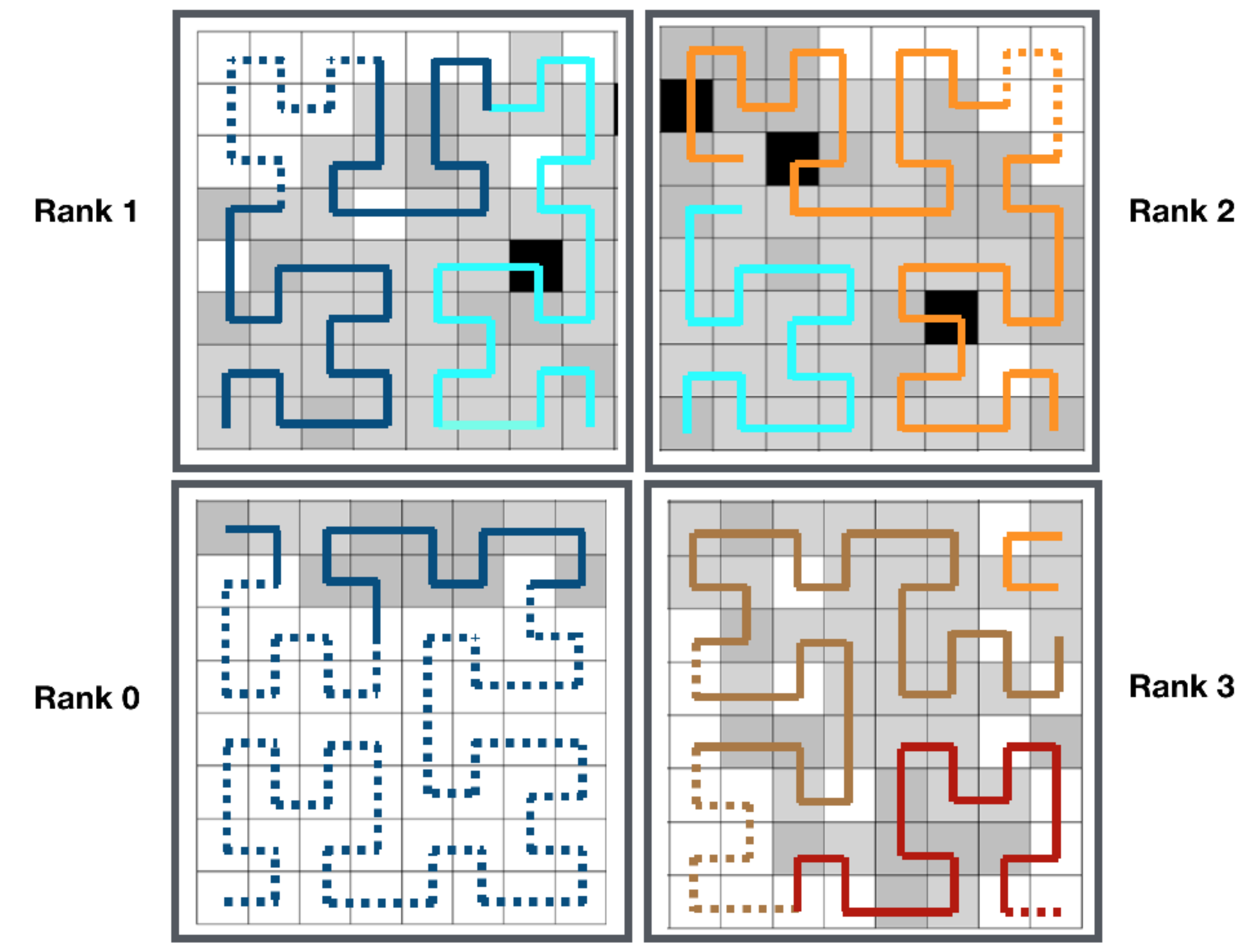}
\caption{Parallel mesh partitioning using SFC.}
\label{fig:sfc}	
\end{figure}

The steps of the SFC-based mesh partitioner are the following: i) a bounding box is defined enclosing the domain; ii) a regular grid is defined 
inside the bounding box; iii) the bins of the regular grid are weighted according to the elements contained on them; iv) the SFC curve is 
used to project the bins to a 1D space; and v) the 1D partitioning problem is solved. 
In the parallel implementation, we divide the bounding box into sub-boxes, and each parallel process performs a local partition within its sub-box. 
Connecting the local partitions, we obtain a coherent partition of the overall mesh. We illustrate this idea in Figure~\ref{fig:sfc}: a parallel SFC is formed 
over a grid of weighted bins, the color of the lines represent the parallel
process to which each bin is assigned, and the gray intensity represents the weight
of the bin. 
A relevant property of our implementation is that the final partition is independent (discarding round-off errors) of the number of processes used to compute it. 
Moreover, the algorithm has no bottleneck in terms of memory or computing requirements. 
A detailed description of this parallel partitioning strategy can be found
in~\cite{BORRELL2018}. In that work we show results for the partition of a  $30$M elements mesh
using $4$K cores of the  BlueWaters supercomputer from the University of Illinois.

\subsection{Load balancing}
\label{sec:loadbalancing}

Heterogeneity, in the mesh elements, in the algorithms or the hardware, complicates the generation of partitions with a uniform work-load distribution. 
Imbalance generates waste of resources since the processes that arrive earlier at the synchronization points remain idle.
The execution traces presented in Section~\ref{sec:perf_char} illustrate the imbalance produced in the airplane simulation considered in this work.

The a priori estimation of the elements weight based on some model is a challenging approach.
Indeed, apart from factors such as the element type or the executing device type, there are also more indirect factors that affect the execution time.
For example, the layout in memory of an element with respect to its neighboring elements determines the cache misses and thus affects its computing cost. 
We could instrument the code to compute accurate weights, but it is hardly possible on an element basis in Alya because it would produce too much overhead.
In this paper, we have designed and implemented a dynamic load balancing strategy by monitoring computing times in a subdomain basis and correcting the partition accordingly. 
Ultimately, we aim to generate partitions to use the heterogeneous \mn{}  \power{} for airplane simulations with unstructured heterogeneous meshes. 
Since the computing power of the GPU is considerably higher than that of the CPU core (see Figure~\ref{fig:gpuassembly_perf}), 
we expect to end up with highly anisotropic partitions with the subdomains assigned to MPI processes running on the GPU having more load.

In this paper, we take advantage of the low latency of the SFC partitioning process to produce corrected partitions according to runtime measurements. 
We carry out this process iteratively. In particular, we specify the correction of the partition through a new input argument added to the SFC partitioner: 
the array $\Lambda[...]$ of dimension $P$. 
When this argument is provided, the target weight for the $i$th subdomain will be 
\begin{EQ}[rcl]
\Gamma_i &=&\lambda_i \frac{W}{P},
\end{EQ}
where $\lambda_i$ is the $i$th component of the array $\Lambda$ and $W=\sum_{i=1}^N{w_i}$ is the sum of the elements 
weight. 
Hence, the mechanism to improve the load 
balancing is not by improving the accuracy of the elements weight but by modifying the target weight per subdomain. 
An obvious requirement is that $\sum_{i=1}^N{\lambda_i} = 1$, so that the weight distributed among the processes is exactly $W$. 

Next, we explain the iterative process to evaluate the correction coefficients $\lambda_i$ that amend the imbalance. First we introduce 
some notation: $t^k_i$ refers to the elapsed time spent by process $i$, in the part of the code under consideration, 
for the $k$th iteration of the balancing algorithm. The superindex $k$ will be used hereafter to refer to the $k$th iteration 
of the balancing algorithm. $\bar{t}^k=\sum_{j=i}^{P}{t_i^k}$ is the average time.  
$T^k_i=\sum_{j=1}^{i}{t^k_j}$ is the sum of the elapsed time of the first $i$ parallel processes. 
$\Upsilon^k_i=\frac{T^k_i}{\bar{t}^k}$ is the normalized sum. And  $\Lambda^k_i=\sum_{j=1}^{i}{\lambda^k_j}$ 
the sum of the first $i$ correction coefficients.

The strategy of the algorithm is to evaluate each SFC splitting point independently. Where by splitting point we refer to  
 subdomains division. For each $i$ we find 
the value $\Lambda^*_i$ such that  $\Upsilon^*_i=i$. So we define the splitting point between the first $i$ subdomains 
and the rest. The balancing process consists in evaluating a simple linear regression (SLR) from the existing 
observations $(\Lambda^k_i,\Upsilon^k_i)$ and cut the regression line with the line $y=i$ to find $\Lambda^{k+1}_i$. Therefore, if 
 the result of the linear regression is $y=\alpha+\beta x$ then
 \begin{EQ}[rcl]
  \Lambda^{k+1}_i &=& \frac{i-\alpha}{\beta}.
 \end{EQ}

 \noindent Finally, the new correction coefficients will be 
  \begin{EQ}[rcl]
 \lambda^{k+1}_i &=& \Lambda^{k+1}_i-\Lambda^{k+1}_{i-1},
  \end{EQ}
 where $\Lambda^{k+1}_0=0$ and $\Lambda^{k+1}_P=P$.

As mentioned earlier, with this method each splitting point is evaluated independently. 
For executions in homogeneous systems, we have observed that $\bar{t}^k$ is practically constant along the iterations (see next subsection). 
In other words, the sum of the time for all processes is independent of how the workload is distributed among them.
In the same way, $\Upsilon_i^k$ is also independent of the definition of the splitting points for the first $i-1$ subdomains. 
Consequently, the convergence of the sequences $(\Lambda^k_i,\Upsilon^k_i)$ is mutually independent. 
Another property of the method is its resilience to outlier measurements since those are smoothed by the linear regression. 
On the other hand, to accelerate the convergence, we use a weighted linear regression (WLR) instead of an SLR, 
the weight is increased by $50\%$ in each iteration to foster the latest observations.

\subsection{Balancing experiments}

In this paper, we use as input for the balancing algorithm the computing time of the elements assembly phase. 
In the pure CPU execution, this kernel represents 75$\%$ of the time step for
the $31.5$M  elements mesh executed using $4$ nodes, 
and 68$\%$ for the  $176$M elements mesh executed using 12 nodes. 
Figure~\ref{fig:trace_phases}, illustrates the dominance of this kernel and the disparity of its execution time across the MPI processes.

The imbalance of an execution is evaluated as 
 \begin{EQ}[rcl]
 I &=& \frac{max(t_k)}{\bar{t}},
\end{EQ}
 where $t_k$ are the input measurements.  
We also refer to the imbalance of the $k$th subdomain as $I_k=\frac{t_k}{\bar{t}}$, what is actually its normalized time.
Another metric that we use is the deviation  $D_k=|t_k-\bar{t}|$. 

\begin{figure}[!h]
  \centering
    \includegraphics[width=0.47\textwidth]{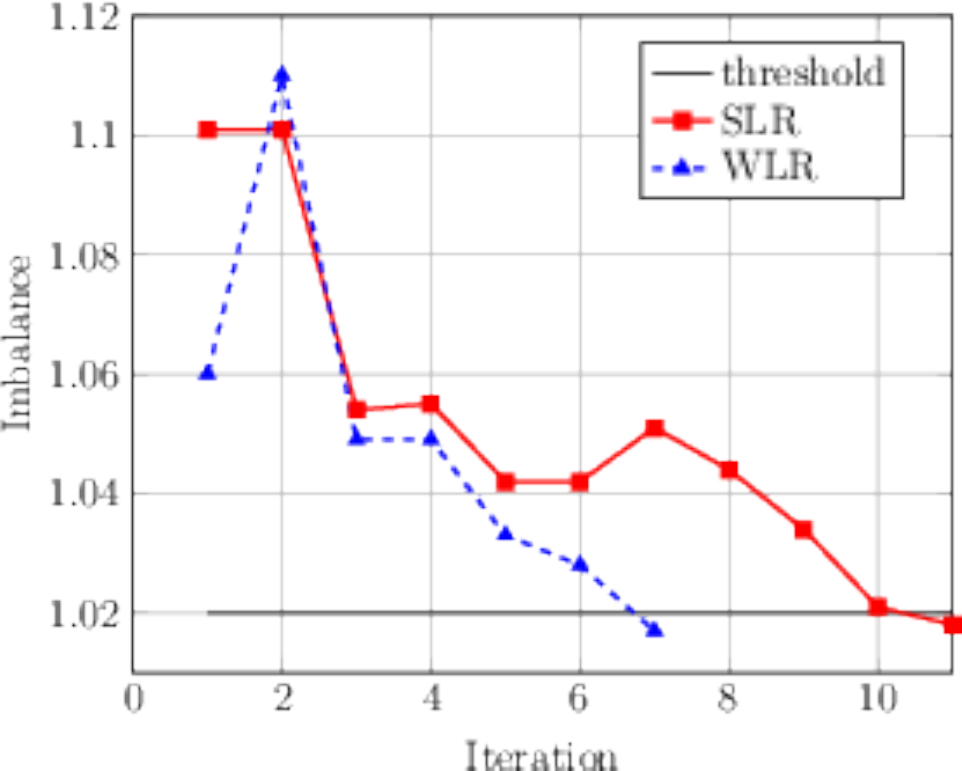}
    \includegraphics[width=0.47\textwidth]{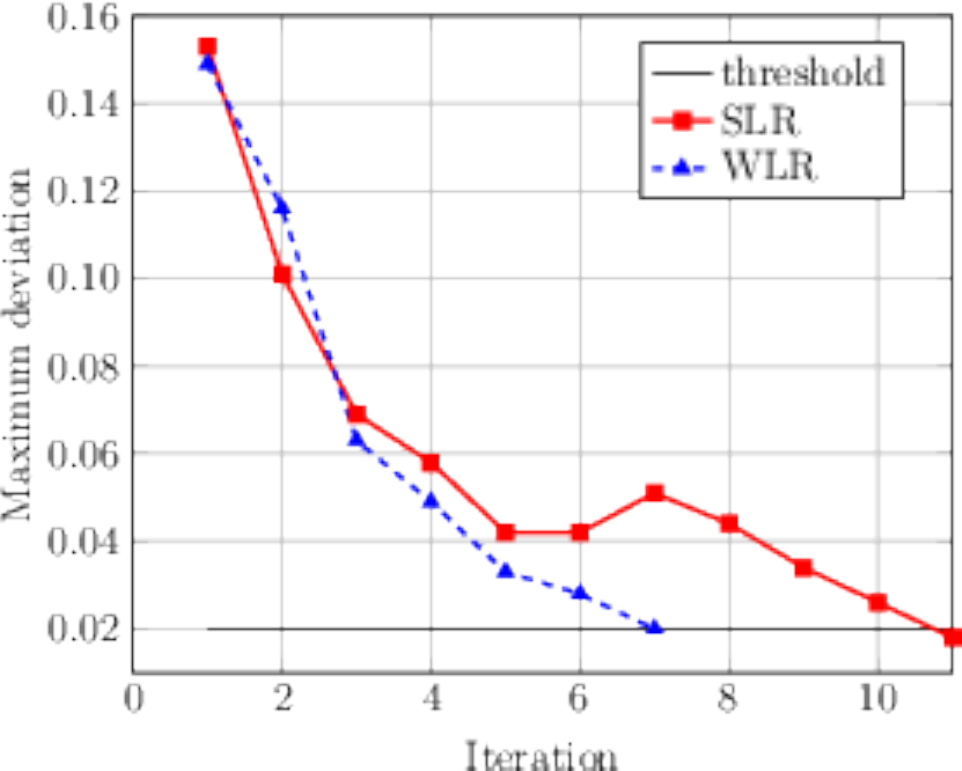}
\caption{Load balancing of the assembly phase for the $31.5$M elements mesh. Comparison of the
SLR and WLR approaches. Left: Imbalance. Right: Max deviation.}
\label{fig:comp}	
\end{figure}

Figure~\ref{fig:comp} shows the performance of the balancing algorithm for the airplane simulation using $8$ CPUs (160 cores) 
of the \power{} for the $31.5$M elements mesh. 
The performance is measured both in terms of imbalance (left) and maximum deviation (right). 
We compare the utilization of the simple linear regression (SLR) versus the weighted linear regression (WLR) within the iterative process. 
The WLR accelerates the convergence, reaching the threshold of $2\%$  in $7$ iterations instead of the $11$ required by the SLR. 
The maximum deviation (right) is smoother. Indeed, the algorithm reduces the deviations, $D_k$, not the imbalance which only depends on the maximum time.

\begin{figure}[!h]
  \centering
    \includegraphics[width=0.8\textwidth]{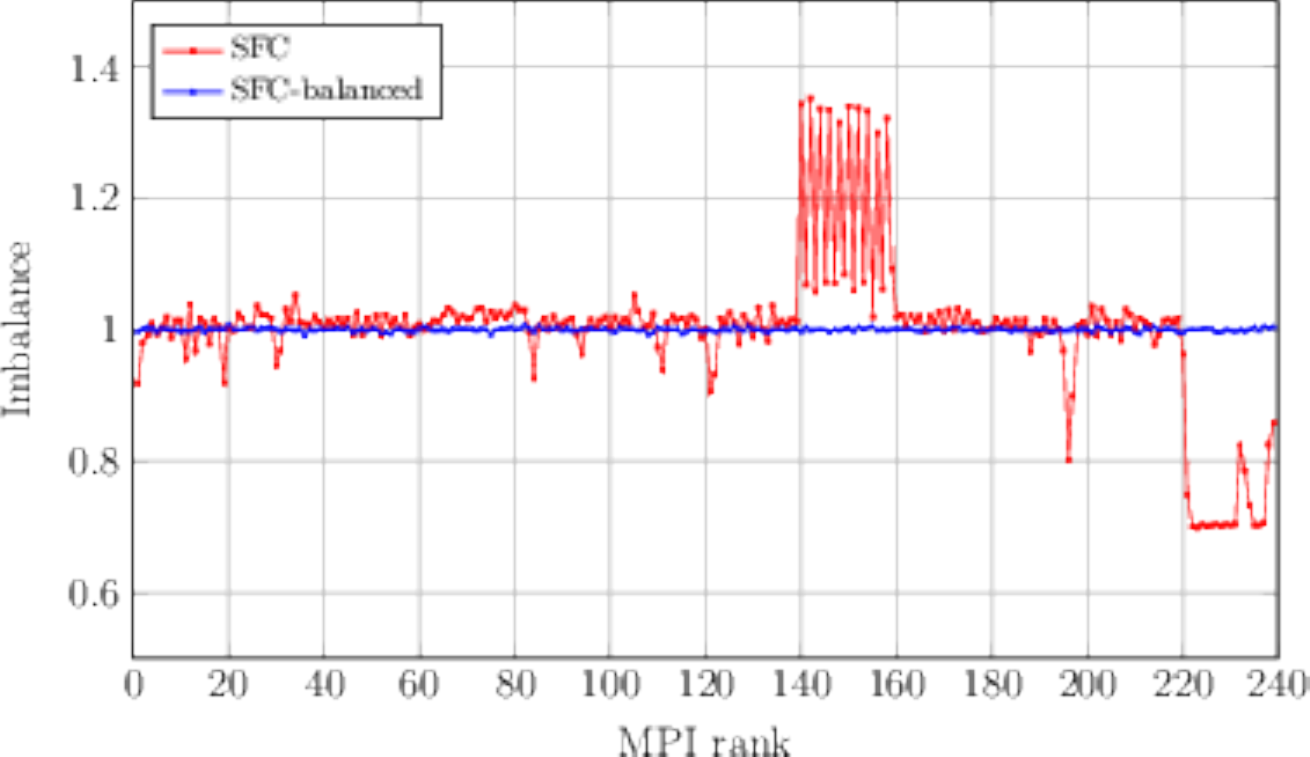}
\caption{Normalized elapsed time ($I_k$) per MPI rank. Assembly phase of the airplane simulation ($176$M elements mesh).}
\label{fig:cpuomps}	
\end{figure}

Now we consider the airplane simulation for the  $176$M element mesh executed in the CPUs of 12 nodes. 
We launch $20$ MPI processes per node, with two OmpSs threads per process, this requires $240$ subdomains.
Figure~\ref{fig:cpuomps} shows the normalized elapsed time $I_k$ of each MPI rank, for the initial and the final execution of the balancing process. 
The imbalance is reduced from $35\%$ down to $0.8\%$. The most significant imbalance occurs for the MPI processes located into two nodes, what
suggests that the performance across the nodes of the \power{} system is not homogeneous.  

\begin{figure}[!h]
  \centering
    \includegraphics[width=0.47\textwidth]{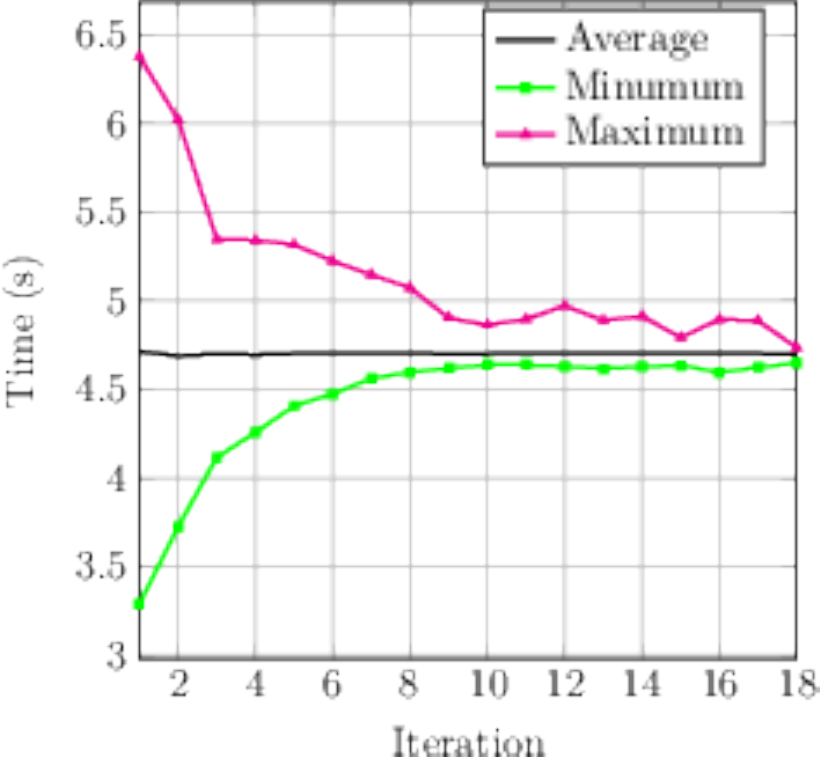}
    \includegraphics[width=0.47\textwidth]{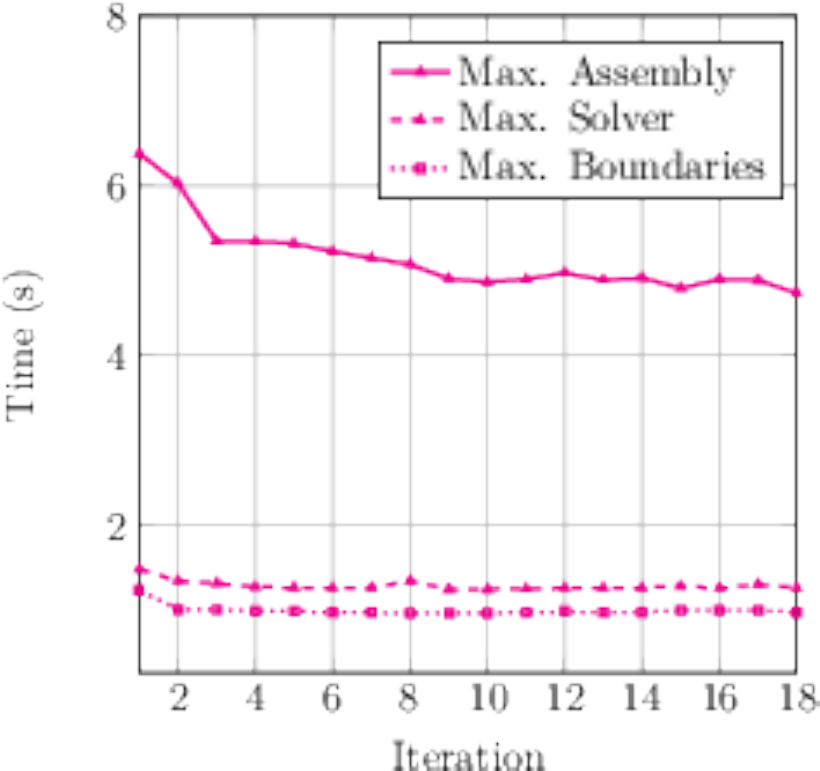}
\caption{Convergence of the balancing process (176M mesh). Left: Evolution of the maximum, minimum and average time for 
the assembly phase. Right: evolution of the different phases of the time step.}
\label{fig:conv}	
\end{figure}

Figure~\ref{fig:conv} (left) shows the evolution of the maximum, minimum and average elapsed time for the assembly phase through the balancing process. 
Both the maximum and the minimum tend to the average. 
The average time $\bar{t}^k$ is almost constant along the iterations, i.e., it is independent of the partition. 
As described in the previous section, this makes the convergence of the splitting points mutually independent. 
We also observe that from the iteration $10$ the convergence stalls. 

\begin{figure}[!h]
  \centering
    \includegraphics[width=0.65\textwidth]{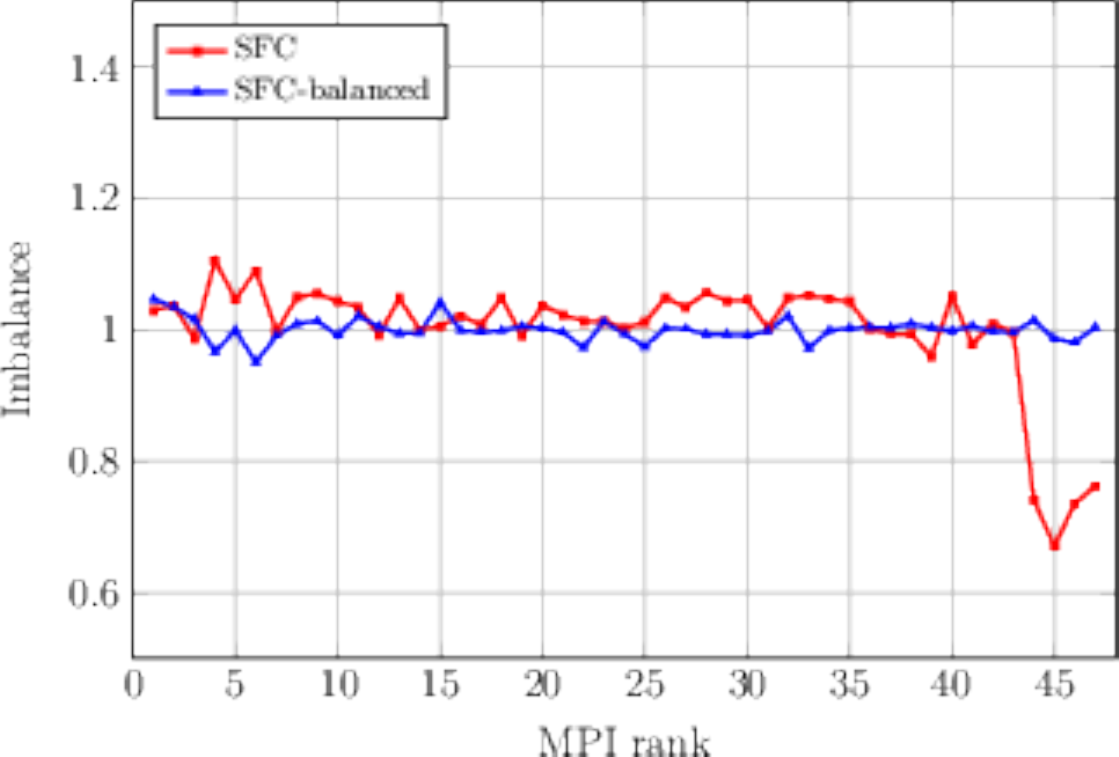}
    \includegraphics[width=0.32\textwidth]{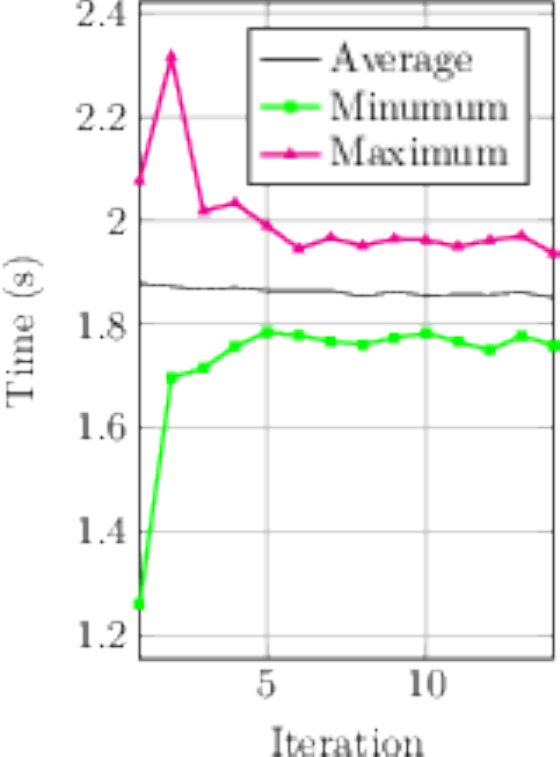}
\caption{Balancing of the assembly phase for the pure GPU execution on $12$ nodes (176M mesh). Left: normalized elapsed time ($I_k$)
per GPU. Right: Evolution of the maximum, minimum and average time.}
\label{fig:gpu}	
\end{figure}

In Figure~\ref{fig:conv} (right), we show the evolution of the elapsed time for the three phases of the time step: 
the element assembly, the boundary assembly, and the linear solver. 
We observe that the element assembly phase dominates the execution and the balancing of this phase does not degrade the rest.
Finally, in Figure~\ref{fig:gpu} we present the same tests, 
i.e., imbalance distribution across MPI processes (left) and convergence of the balancing process (right), for the pure GPU execution. 
The same $12$ nodes are involved. Therefore a total of $48$ Volta V100 GPUs are engaged. 
Qualitatively, results are similar to those obtained for the CPUs. 
The final time of the GPUs based execution (1.9s) is $2.5\times$ lower than the CPUs one (4.7s), this is coherent with the results shown in Section~\ref{sec:cpu}.

\section{Co-execution}
\label{sec:coex}
 
At each time step, the numerical method for the solution of turbulent flows has three main phases, see Algorithm~\ref{alg:kernels}: 
the element and boundary assemblies, carried out at each Runge-Kutta loop, and then the solution of the pressure equation. 
In the pure CPU execution, for the 176M elements mesh, the two assembly phases represent roughly
$80\%$ of the computing time. 
In this paper we have developed a co-execution strategy for the assembly loops
while the linear solver is exclusively executed on the GPUs. In the following subsection, we derive a simplified model of the co-execution
efficiency to predict the gain  with respect to the
pure CPU or pure
GPU executions. We then explain in next subsection the multi-code strategy adopted and
finally present several computing experiments.

\subsection{Theoretical co-execution efficiency}
\label{sec:theoretical}

Heterogeneous architectures pose the challenge of computational efficiency as, in general, one of the devices is used at the expense of the other. 
The co-execution model proposed in this paper aims at executing the same kernel (element assembly, see Section \ref{sec:app}) concurrently on CPUs and GPUs,
with well-balanced workload (Section \ref{sec:loadbalancing}) so that computational efficiency is enhanced with respect to single device execution. 
In the following we will devise a simple performance model to assess the potential of such a co-execution model,
assuming a perfect load balance among the available devices.

Let us consider a computational kernel (herein the elements loop) to be executed,
using one of the available devices, CPU core or
GPU, or both of them in a co-execution mode. 
Let $n_{core}$ and $n_{gpu}$ be the numbers of available cores and GPUs, respectively, and define the speedup $s$
of the GPU with respect to a single core. The normalization of the speedup with respect to a single core
timing has been chosen for the sake of generality. Let us finally note that the speedup is in practice impossible to predict as the
time for element assembly not only depends on the type of element, but also on the subdomain load,
that is the total number of elements to assemble.
Using these definitions, the total available (weighted) resources are $(n_{core}+s\times n_{gpu})$. 

Let $r:=n_{gpu}/n_{core}$ be the ratio of GPUs per core which, in the case of the present \power{}
configuration is $1/10$.
Let us now define three computational efficiencies, measuring the ratio of used resources over available
resources. $\eff_{gpu}$ is the efficiency one would obtain by using exclusively the
GPUs; $\eff_{core}$ is the efficiency using only the cores of the CPUs; finally, we
define the co-execution efficiency as $\eff_{coex}$. We now assume a perfectly load
balanced computational kernel and compute the three efficiencies. For an execution
using only the GPUs:

\begin{EQA}[rcl]
\eff_{gpu} &=& \frac{s \times n_{gpu}}{n_{core}+s \times n_{gpu}}, \\
&=& \frac{s \times r}{1+s\times r }. \label{eq:effgpu}
\end{EQA}
When using only the cores, we have
\begin{EQA}[rcl]
\eff_{core} &=& \frac{n_{core}}{n_{core}+s \times n_{gpu}}, \\
&=& \frac{1}{ 1 + s \times \displaystyle{r} }. \label{eq:effcpu}
\end{EQA}
Eventually, when considering a balanced co-execution, we consider two different scenarios.
For the first option, only one MPI process is assigned to each
GPU and the corresponding core remains idle. For the second option, when using OmpSs, two cores are
lost for each MPI process assigned to the GPU.
Thus, in the first case, $n_{gpu}$ cores no longer participate 
to the co-execution so that the used resources are $(n_{core}-n_{gpu}+s \times n_{gpu})$, implying that
\begin{EQA}[rcl]
\eff_{coex,1} &=& \frac{n_{core}-n_{gpu}+s \times n_{gpu}}{n_{core}+s \times n_{gpu}}, \\
&=& \frac{1+(s-1) \times r}{ 1 + s \times r }.\label{eq:coex1}
\end{EQA}
In the second case, $2n_{gpu}$ cores no longer participate 
to the co-execution so that the used resources are $(n_{core}-2n_{gpu}+s \times n_{gpu})$, implying that
\begin{EQA}[rcl]
\eff_{coex,2} &=& \frac{n_{core}-2n_{gpu}+s \times n_{gpu}}{n_{core}+s \times n_{gpu}}, \\
&=& \frac{1+(s-2) \times r}{ 1 + s \times r }. \label{eq:coex2}
\end{EQA}

\begin{figure}[!h]
  \centering
  \includegraphics[width=0.47\textwidth]{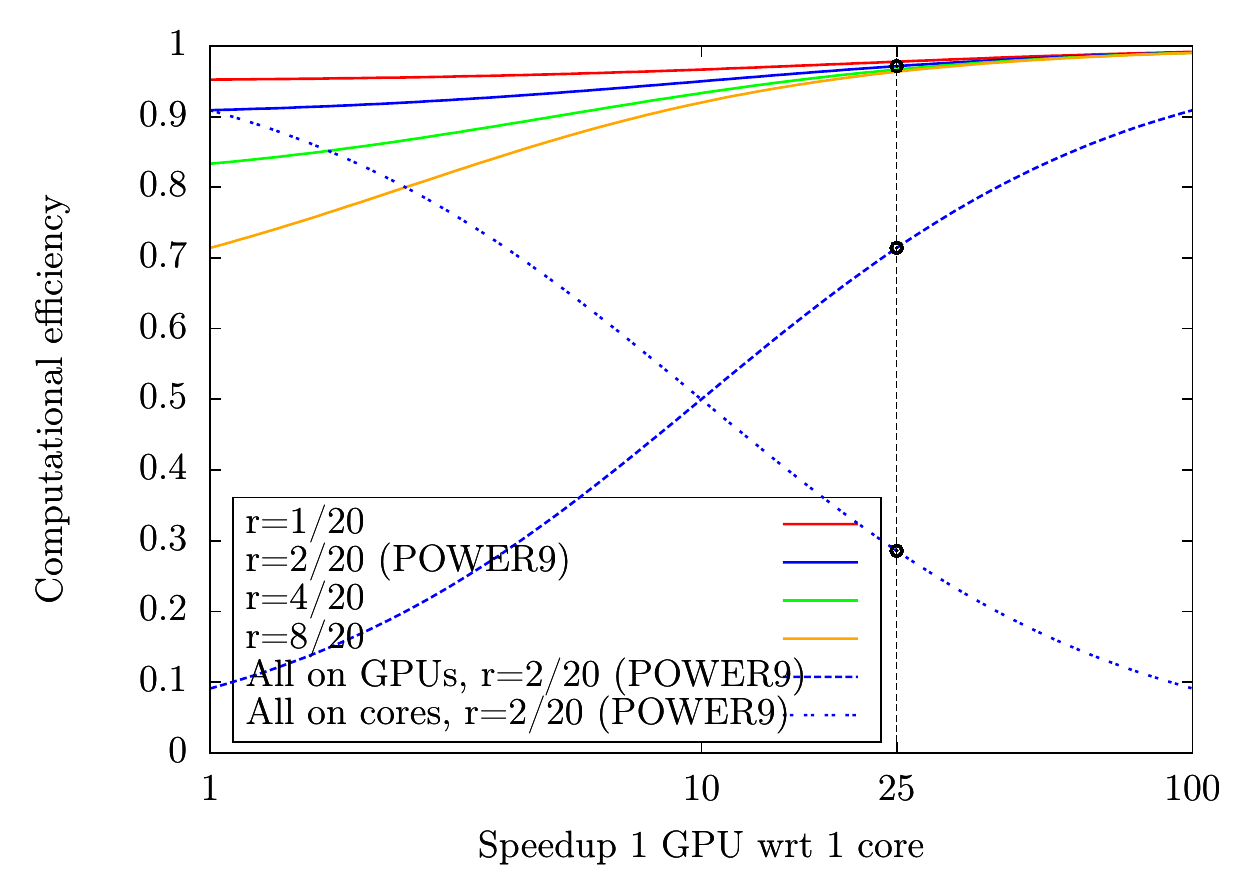}
  \includegraphics[width=0.47\textwidth]{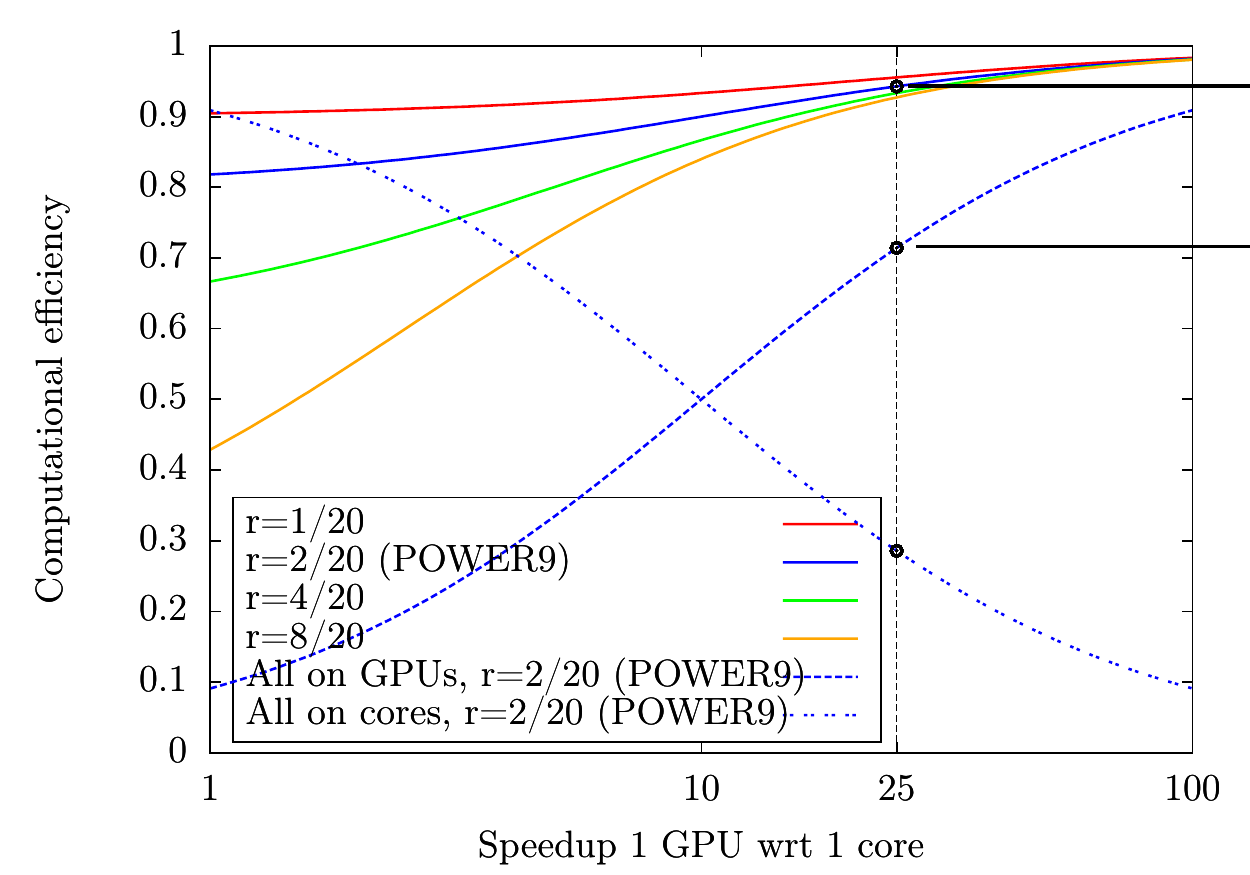}
  \caption{Efficiency of co-execution compared to executions on cores or GPUs, exclusively.
  The dots show the actual \power{} configuration of \mn{} at a speedup of 25. Left: one core
  per GPU. Right: two cores per GPU.}
  \label{fig:theocoex}
\end{figure}

Figure~\ref{fig:theocoex} shows the behavior of the previously defined efficiencies.
On the left hand-side, assigning one core per GPU (Equations \eqref{eq:effgpu}, \eqref{eq:effcpu} and \eqref{eq:coex1}) and on the right-hand side, assigning two cores
per GPU (Equations \eqref{eq:effgpu}, \eqref{eq:effcpu} and \eqref{eq:coex2}).
For both scenarios, in the case of the exclusive use of cores or GPUs, we have considered the configuration of the \mn{}  \power{} ($r=2/20$) . 
On the one hand, the efficiency obtained by using only cores decreases when the speedup, $s$, increases.
The efficiency reaches as little as $10\%$ for a speedup of 100, under which circumstances the cores do not provide relevant resources.
On the other hand, the efficiency obtained when using only GPUs increases with
growing speedup, reaching $90\%$ with a speedup of 100.

In the case of co-execution, different numbers of GPUs to cores ratios $r$ have been considered. 
With a small speedup, increasing this ratio $r$ reduces greatly the efficiency; in the case of a unit speedup, a GPU is equivalent to a core, and
core resources are idle at the expense of using GPUs, thus decreasing the efficiency. 
Assigning one core per GPU (see left-hand side of Figure~\ref{fig:theocoex}), and in the case of having as much GPUs as cores and a speedup of 1, the efficiency
would reach $50\%$, as half of the
resources would remain unused. 
We observe that for speedups over 20, the efficiency is quickly overpassing
$90\%$, showing the gain brought by the co-execution.

Finally, let us investigate the specific case of our airplane simulation.
Figure \ref{fig:gpuassembly_spd} shows that, in average, the speedup obtained using 4 GPUs with
respect to 40 cores is approximately $2.5$. Using the present definition, this corresponds to a speedup $s=25$,
(one GPU is as fast as 25 cores).
The dots in the figure depict the efficiencies at this speedup obtained with the actual \power{} configuration of \mn{}.
We notice that using only cores, the efficiency is $30\%$. Using exclusively GPUs, the
efficiency is around $70\%$. Finally, co-execution is just below $100\%$, thus enabling to increase the
efficiency by almost $30\%$ with respect to using GPUs exclusively.
In fact, the efficiency of co-execution never reaches $100\%$, as the cores
assigned to the GPUs do not participate in the computation. We also observe that adding more GPUs
would not affect significantly the efficiency in the specific case of speedup 25.
Finally, when comparing the use of one or two cores per GPU (left and right figures, respectively), we observe that
the impact of co-execution is more limited in the second case, as twice more cores remain unused.

\subsection{Multi-code approach}

Our approach for the co-execution of the assembly phase on heterogeneous systems is basically a standard domain decomposition; i.e. 
the same kernel is executed simultaneously by different MPI processes associated each one to a subdomain.
The load balancing algorithm for co-execution is also the same that 
we use for executions on homogeneous systems; i.e. the partition is adapted from runtime measurements through the correction coefficients. 

The only difference of the co-execution is that we generate a specific executable for each device. In other words, the executable run 
by each MPI process depends on the device to which it is mapped. That is what we refer to as a multi-code approach. The \mn{} \power{} 
relies on the SLURM queuing system, and this mapping of MPI ranks and executables can be managed with the \code{--multi-prog} flag of 
the \code{srun} command. 
The only existing requirement for the parallel execution to work is that all the engaged
executables are linked to the same
MPI library. Indeed, the unique interaction between executables are the MPI instances for data exchange and synchronization.

In agreement with the best results raised in Section~\ref{sec:cpu}, we use the
GFortran compiler to generate the CPU executable, and 
PGIF90, that supports OpenACC, for the GPU. The compilation parameter \code{PACK\_SIZE}, that determines the granularity
of the vectorization, is also optimized for each device. In particular, coherently with the tests presented in Sections~\ref{sec:cpu} 
and~\ref{sec:gpu}, we set it to $32$ and to $190K$ in the CPU and GPU compilations, respectively.

\subsection{Co-execution experiments for the Airplane simulation}

\begin{figure}[!h]
  \centering
    \includegraphics[width=0.80\textwidth]{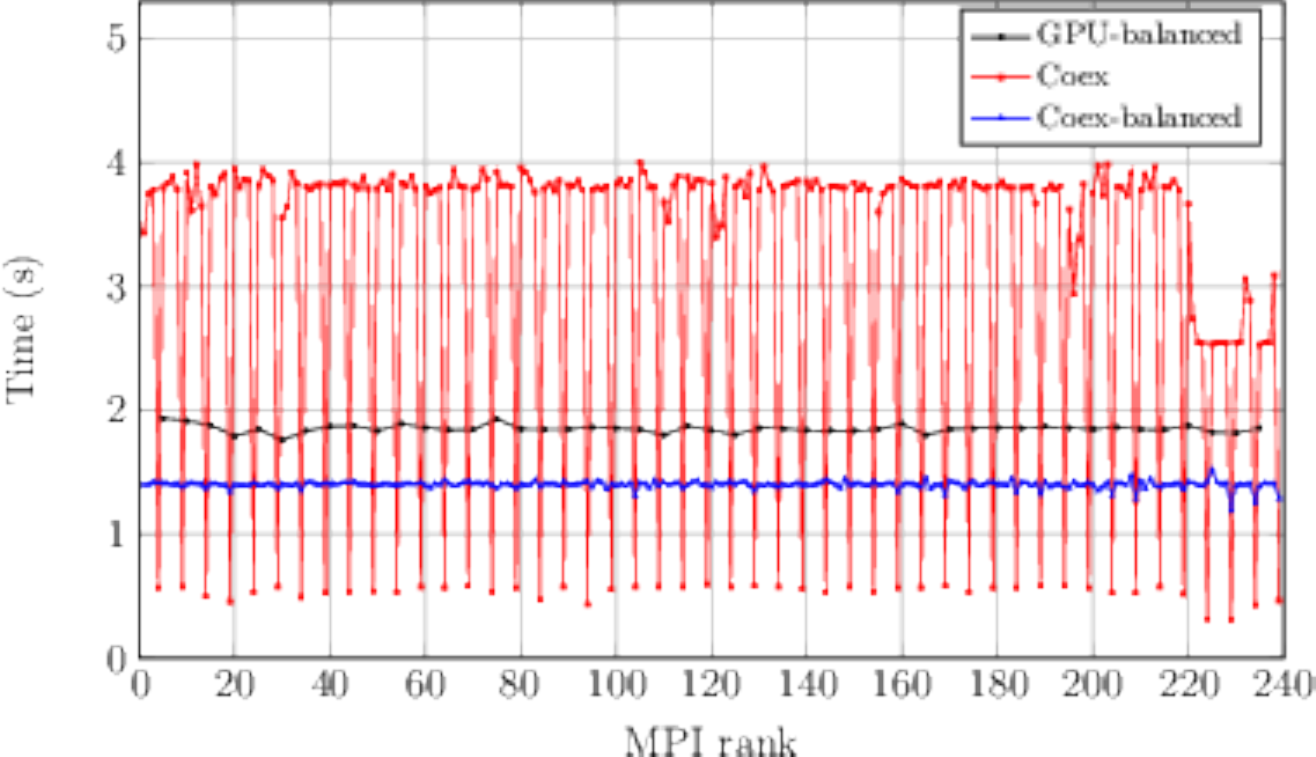}
\caption{Balancing of the assembly phase for the  co-execution on $12$ nodes (176M mesh), comparison with the pure GPU execution.}
\label{fig:coex}
\end{figure}

We consider now the co-execution of the $176$M elements mesh on $12$ nodes. Figure~\ref{fig:coex} shows the elapsed time
 of the elements assembly phase per MPI-process on the initial and final iteration of the balancing process; together with the 
 result corresponding to the balanced partition for the pure GPU execution.

In the first iteration of the co-execution, the same load is assigned to each MPI process, whether it runs on a CPU or a GPU
device. As a result we see a great disparity between the elapsed times depending on the type of device where the associated MPI
is assigned, but also a disparity between the MPIs assigned to the same type of device. In the final iteration we observe that
both unbalances are canceled, and a rather good balance (exactly $94\%$) is obtained. In particular, in average the subdomains
executed on the GPUs end up with roughly $10\times$ more elements than those running on the CPU cores. Having two OmpSs threads per
subdomains, this corresponds to a speedup of $s=20$ as defined in Section \ref{sec:theoretical}. 
To summarize the configuration of the case, we have per node:
\begin{itemize}
\item 16 MPI subdomains solved on CPU-cores, with 2 OmpSs threads per MPI subdomain.
\item 4 MPI subdomains executing on GPUs (this results in 8 idle CPU-cores on the host).
\item Power9 configuration: $r := n_{gpu} / n_{core} = 1/10$.
\item Measured speedup for the assembly $s=20$.
\end{itemize}
With these figures, and using Equation \eqref{eq:coex2}, we obtain that
$\eff_{coex,2}= 93 \%$, whereas the efficiency using exclusively the GPU is $\eff_{gpu} = 67$,
using Equation \eqref{eq:effgpu}.
Recalling that the computing time is inversely proportional to the efficiency, this model  
predicts a $28\%$ reduction of the elapsed time at using the co-execution $vs$ the pure GPU execution, 
in the experiments presented in Figure~\ref{fig:coex} the difference measured is $23\%$.

In practice, this means that by engaging the host on the elements assembly calculations, we obtain the same performance boost as if we 
added an additional GPU to each node. Leaving this performance on the table would clearly be a suboptimal use of the machine.

\begin{figure}[!h]
  \centering
    \includegraphics[width=0.40\textwidth]{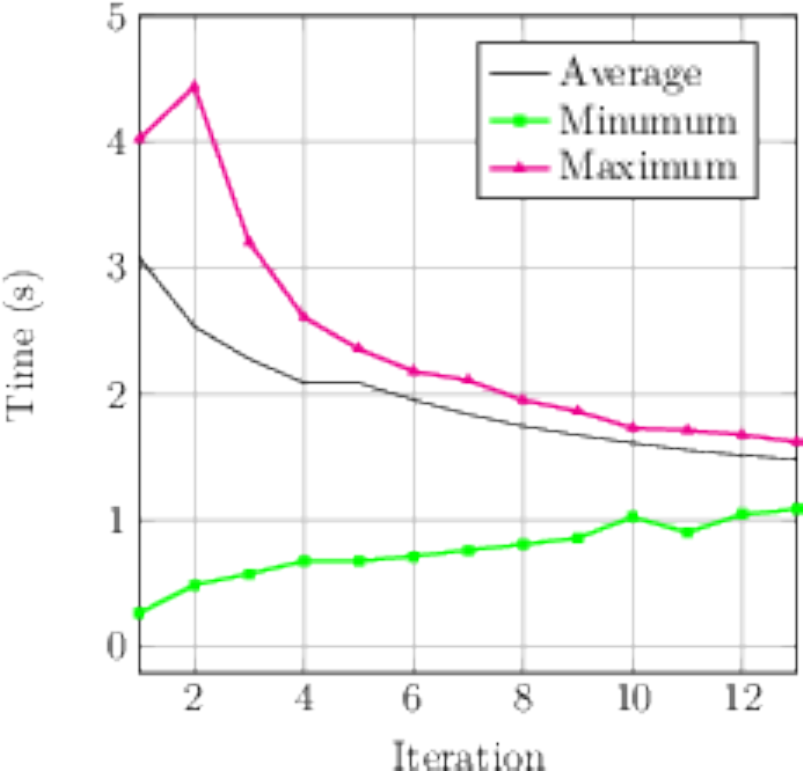}
    \includegraphics[width=0.40\textwidth]{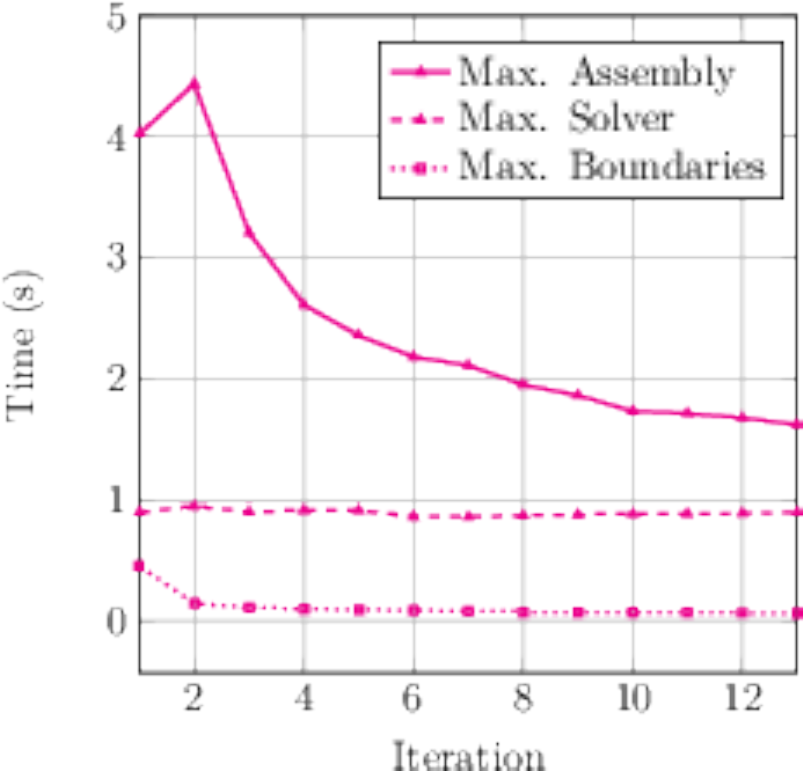}
\caption{Convergence of the balancing process for the co-execution (176M mesh). Left: Evolution of the maximum, minimum and average 
time for the assembly phase. Right: evolution of the different phases of the time step.}
\label{fig:coex2}
  \end{figure}

Figure~\ref{fig:coex2} (left) shows the convergence of the maximum, minimum and average time along the iterative balancing process. 
Notice that, unlike the homogeneous case, the average time is not constant but converges asymptotically. This is explained by the 
fact that when moving elements from the CPU to the GPU there is a significant drop on its execution time that decreases the average.

In Figure~\ref{fig:coex2} (right) we show the evolution of the elapsed time for the three phases of the time step. The partition 
is adapted to the assembly phase, however the elapsed time for the other two does not increase but even decreases for the boundaries
assembly phase. Note that in the solver phase all the MPI processes run on the GPU. In particular, five MPI processes execute the 
solver on each GPU: four that had run the assembly  on the CPU cores and one on the same GPU. On the balancing process the workload 
of the GPU processes increases (the final ratio is about 1:10).  However,  the sum of the load of each group of  five processes that 
end up running the solver on the same GPU does not change much along this process. As a consequence, the solver time keeps almost 
constant, being independent of the load distribution between the processes assigned to each GPU.

%% file: sections/07_conclusions.tex
\section{Conclusions}
\label{sec:conclus}

In this paper we have described the approach implemented in Alya for the
efficient use of the accelerated \power{} 
installed at BSC. This cluster is based on a similar architecture than the Summit and Sierra supercomputers, ranked in the 
first two positions of the TOP500 list (November 2019). We have assessed the performance of Alya using up to 40 nodes for 
simulations of airplane aerodynamics.

We have outlined the numerical formulation implemented in Alya for the  simulation of turbulent flows, 
along with a more detailed description of its multilevel parallelization strategy.
In Alya, the distributed and shared memory approaches are complemented, at the last level,
with vectorization on the CPU and stream processing on the GPU. 
A remarkable feature is that the same data-structuring, and de facto the same code, is used in both cases 
although with a different level of granularity.

We have presented a detailed analysis of the performance of the code in both the POWER9 AC922 CPU and the 
NVIDIA Volta V100 GPU. We have studied the CPU performance considering different compilers and shared
memory configurations.
Regarding the GPU, we have  presented the data layout as well as the concurrent execution 
model used to attain the maximum performance. With the final implementation, for the assembly phase we have achieved a speedup 
of $3.48\times$ vs the CPU execution for unstructured heterogeneous meshes. Such results match with previous works of CFD simulations using 
unstructured meshes.

Finally, we have focused on the efficient exploitation of the whole node through heterogeneous computing.
We have presented a novel approach based on a standard domain decomposition to parallelize a multi-code execution, 
together with a dynamic load balancing mechanism to adapt the underlying mesh partition to the performance of each device. 
The goal is to adjust the partition to the average performance of each device, eliminating the structural imbalance.

The balancing algorithm is thus the intrinsic tool that we use for co-execution on heterogeneous systems. 
To adapt the partition to the measured imbalance, we have added subdomain correction coefficients to an in-house SFC-based 
mesh partitioner.
A linear regression is used to define the splitting points in the 1D space generated by the SFC projection. 
This approach makes the method robust to outlier measurements, converging to well-balanced partitions in few iterations. 
For example, for a heterogeneous execution of the full airplane simulation using a 176M elements mesh, we achieve a load 
balance of $94\%$ in $14$ iterations. 
 
Finally, we have applied our co-execution strategy to the assembly phase of the
time step, keeping the execution of the linear solver entirely 
on the GPUs. We observe a reduction of the assembly time by 23$\%$ compared to the pure-GPU execution. 
In practice, this means that by properly engaging the CPU on the calculations, the performance boost achieved is equivalent to 
having an additional GPU per node.